\documentclass[useAMS,usenatbib,usegraphicx]{mn2e}

\def\la{\raise.5ex\hbox{$<$}\kern-.8em\lower 1mm\hbox{$\sim$}}
\def\ma{\raise.5ex\hbox{$>$}\kern-.8em\lower 1mm\hbox{$\sim$}}

\def\msol{M$_{\odot}$ }
\def\kms{$\rm km\, s^{-1}$}
\def\cm3{$\rm cm^{-3}$}
\def\Ts{$\rm T_{*}$}
\def\Vs{$\rm V_{s}$}
\def\n0{$\rm n_{0}$}
\def\B0{$\rm B_{0}$}

\def\erg{$\rm erg\, cm^{-2}\, s^{-1}$}
\def\mum{$\mu$m}

\def\mum{$\mu$m}

\def\L12{L$_{12\mu m}$~}
\def\F12{F$_{12\mu m}$~}

\title[The SED of D-type symbiotic stars]{The spectral energy distribution of D-type symbiotic stars:
the role of dust shells}

\author[R.Angeloni, M.Contini, S.Ciroi, P.Rafanelli]{R. Angeloni\thanks{\textit{Current address}: Departamento de Astronom\'{i}a y Astrof\'{i}sica, Pontificia Universidad Cat\'olica de Chile, Av. Vicu\~na 
Mackenna 4860, 782-0436 Macul, Santiago, Chile}$^{1,2}$, M. Contini$^{2,1}$, S. Ciroi$^{1}$ and P. Rafanelli$^{1}$ \\
$^{1}$Dipartimento di Astronomia, Universit$\grave{a}$ di Padova, Vicolo dell'Osservatorio 2, I-35122 Padova, Italy\\
$^{2}$School of Physics and Astronomy, Tel-Aviv University, Tel-Aviv, 69978 Israel\\
}
\begin{document}

\date{Accepted . Received ; in original form }

\pagerange{\pageref{firstpage}--\pageref{lastpage}} \pubyear{2002}

\maketitle

\label{firstpage}

\begin{abstract}
We have collected  continuum data
of  a sample of D-type symbiotic stars.
By modelling their spectral energy distribution in a colliding-wind theoretical scenario
we have found  the common characteristics to all the systems:
1) at least two dust shells are clearly present, one  at  $\sim$ 1000 K and the other at
$\sim$ 400 K; they dominate the emission in the IR;
2) the radio data are explained by  thermal self-absorbed emission from the reverse
shock between the stars; while 3)  the data in  the  long wavelength tail
come from the expanding shock outwards the system;
4) in some  symbiotic stars, the contribution from the WD in the UV is directly seen. Finally, 5) for some objects soft X-ray emitted by bremsstrahlung  downstream of the reverse-shock 
between the stars are predicted.
The results thus confirm the validity of the colliding wind model
and the important role of the shocks.
The comparison of the fluxes calculated at the nebula with those observed
at Earth   reveals the  distribution throughout the system
of the
different components, in particular the  nebulae and the dust shells.
The correlation of  shell radii with the orbital period shows that larger radii
are found at larger periods. Moreover, the temperatures of the dust shells
regarding the sample are  found  at $\sim$ 1000 K and $\leq$ 400 K, while, in the case of
late giants, they spread
more uniformly throughout the same range. 
\end{abstract}


   \maketitle

\section{Introduction}

\begin{table*}
\centering \caption{The sample in the literature. \label{tab:lit}}
\begin{footnotesize}
\begin{tabular}{c c c c c c c c c}\\
\hline \hline 
Object & Coordinates & distance & $P_{orb}$ & $P_{Mira}^\textit{a}$ & \.{M}$_{Mira}$ & $T_{Mira}$ & $T_{dust}$ \\
& $\alpha \; [J2000] \; \delta$ &[kpc] & [years]& [days]& [M$_{\odot}$/yr] & [K] & [K] \\
\hline 
BI Cru & 12:23:27.4 $\;$ -62:38:16.5 & 2 & - & 280 & $3 \, 10^{-6}$& - & 1300\\
SS73 38 & 12:51:26.2 $\;$ -64:59:58.8 &4.8& - & 463 & $2.5 \, 10^{-6}$ & - & -\\
V835 Cen & 14:14:09 $\;$ -63:25:45.3 & 2.8 & - & 450 & - & 2250 & 1000\\
H1-36 & 17:49:48.2 $\;$ -37:01:27.0 & 4.5 & $>$95 & 450-500 & $10^{-5} d^{3/2}$ & 2500 & 700 \\
HM Sge & 19:41:57.1 $\;$ 16:44:39.6 & 2.3 & 90$\pm$20 & 527 & $\approx10^{-5}$ & 3000 & 1400\\
V1016 Cyg & 19:57:05 $\;$ 39:49:36.0 & 2.93$\pm$0.75 & 80$\pm$25 & 474 & $\geq 3 \, 10^{-7}$ & 2450$\pm$150 & 600 \\
RR Tel & 20:04:18.52 $\;$ -55:43:33.4 & 2.5 &-& 387 & $1.6\, 10^{-6}$ & 2300 & 1000 \\
V627 Cas & 22:57:42 $\;$ 58:49:14.0 & $<$0.8 & - & 466 & - & 2650$\pm$50 & 1000$\pm$50 \\
R Aqr & 23:43:49.4 $\;$ -15:17:40.3 & 0.2 & 44 & 395 & $10^{-6}$ & 2800 & 1000 \\
\hline
\end{tabular}
\begin{scriptsize}
\flushleft
References: \textit{a}: Belczy{\' n}ski et al. (2000); BI Cru: Schwarz \& Corradi (1992), Contini et al. (2008c); SS73 38: Munari (1992), Gromadzki et al. (2009); V835 Cen: Whitelock (1987), Feast et al. (1983); H1-36: Allen (1983), Angeloni et al. (2007b); HM Sge: Whitelock (1987), Richards et al. (1999), Bogdanov \& Taranova (2001), Schild et al. (2001); V1016 Cyg: Parimucha (2003), Schild \& Schmid (1996), Lee et al. (2003), Taranova \& Shenavrin (2000); RR Tel: Feast et al. (1983), Kotnik-Karuza et al. (2002), Gromadzki et al. (2009); V627 Cas: Bergner et al. (1988); R Aqr: Gromadzki et al. (2009); Hollis et al. (1997), Contini \& Formiggini (2003), Dougherty et al. (1995).
\end{scriptsize}
\end{footnotesize}
\end{table*}

{\it The word "symbiotic" was introduced by Merrill in 1944
and this term is currently used for the category of variable stars with composite
 spectrum. The  main spectral features of these objects are: 1) the presence 
of a red continuum typical of a cool star , 2) the rich emission line spectrum,
and 3) the UV excess. [...]
In addition to the peculiar spectrum, the very irregular photometric and
spectroscopic variability is the major feature of symbiotic stars}.
So Viotti in 1993 introduced the symbiotic phenomenon.

Since then, a large number of symbiotic stars (hereafter SSs) was observed. 
In 1995 we started the analysis of SSs by interpreting the line spectra of RS Ophiuci
(Contini et al. 1995)  by means of shock-dominated models. 
At that time, it  was established that both the WD and red giant stars loose winds which collide within and outwards
the system (Nussbaumer 2000 and references therein).
Therefore, to model the SS systems on a large scale, we adopted the colliding-wind scenario of Girard \& Willson (1987) which leads to different shock fronts throughout the SS.
The nebulae downstream are  illuminated by the flux from the  stars, therefore, a composite model 
(shock + photoionization) was needed, and successfully applied to HM Sge (Formiggini, Contini, Leibowitz 1995). 

 In certain objects accretion disks are formed  at certain epochs. They blow up during the outburst
of the hot component star. The presence of a  disk  is revealed by the jets that were observed and modelled, for instance,
in BI Cru (Contini et al 2009b), He2-104 (Contini \& Formiggini 2001), R Aqr (Contini \& Formiggini 2003),  CH Cyg
(Contini et al. 2009c), etc. The line ratios observed in the  jet regions  show 
lower densities ($\sim$ 200 \cm3)  and velocities (FWHM  of 150-200 \kms) than those
 corresponding to the nebulae created by the wind collision ($\geq$ 10$^7$ \cm3 and $\leq$ 1000 \kms, respectively). 
Therefore, also the continuum flux intensity is lower and  does not significantly affect the SED.

We realized that the results of the line spectrum analysis  of SSs
should be used to constrain the spectral energy distribution (SED) of the continuum.
The observations in the different wavelength ranges provided sufficiently reliable   data.
Therefore, not only the red giant star and some times even the WD  could be recognized 
throughout the SED, but today the plural nature of the emitting nebulae  
and of the dust shells  within the SS systems is well established (Contini 1997;  
Contini \& Formiggini 2001, 2003; Angeloni et al. 2007a, hereafter Paper I; Angeloni et al. 2007b,c and references therein; Contini et al.
2009a,b,c and references therein).

 Moreover, it is difficult to explain the observed
IR SED of dusty SSs by a single temperature, and nowadays theoretical models and observations
too have been confirming that several dust temperatures should be combined in order to
reproduce the NIR-MIR data (Anandarao et al. 1988, Schild et al. 2001, Paper I). This will be definitively demonstrated  by   further data coming from the
MIR-FIR windows that ISO, Spitzer, and AKARI have opened.

The  present analysis  focuses on a sample of D-type (i.e. dusty) SSs, 
producing a detailed picture of each system.
The SS continua are   analysed in a  quantitative  way
mainly through the light-curve variability,  through  the cool and hot
star spectra and through the emission line spectra in the optical and UV range.
In particular, by modelling the data in the continuum, we could clearly recognise
in  each  object the relative  contribution of the dust shells and of the  nebulae.
The shells are ejected as a consequence of pulsations of the Mira, while the 
 nebulae  appear  downstream of the shock fronts created by the collision of the
 star winds. 

As mentioned above, the calculation of the spectra  must account consistently for shocks and 
photoionization; therefore, we use for the calculation of the spectra the code 
SUMA\footnote{http://wise-obs.tau.ac.il/$\sim$marcel/suma/index.htm},
which simulates the physical conditions of an emitting gaseous nebula under the coupled effect of
 photoionization
from an external source and shocks, and in which line and continuum emission from gas are calculated 
consistently with
dust reprocessed radiation as well as with grain heating and sputtering.
SUMA  has been  applied to interpret several SSs, as extensively reported in previous papers
(e.g. Angeloni et al. 2007b; Contini et al. 2009b and references therein), as well as nova stars (V1974, Contini et al. 1997; T Pyx,
Contini \& Prialnik 1997) and supernova remnants (e.g. Kepler's SNR, Contini 2004).

We present the  D-type SS sample  in Table 1.
We gathered for each object the observational data from the literature
and the  results  obtained by modelling the continuum
SED,  which were constrained by the analysis of the line spectra
in each object. 

Our  aim is to characterize the SED of D-type SS; therefore we chose 
our targets on the basis of the best  data available,
in order to create a  SED on a large wavelength range for each object. 
This has allowed to better constrain several physical parameters essential for a
meaningful and comprehensive interpretation, taking  into account some 
issues, e.g.  the effects of intrinsic flux variability on our modelling method. 
In this way, we are able to compare the  features common to the dusty SSs.
Moreover  the correlations of the characteristic physical properties
such as luminosity, orbital period, dust shell sizes, etc.,  is  discussed
on the basis of reliable parameters.

In Sect. 2 the data are presented and discussed in the light of the photometric and spectroscopic variability.
Sect. 3 deals with theoretical aspects: the colliding wind scenario, the calculation
code, and the modelling method. In Sect. 4 the SEDs  are analysed. The results are shown for the single
SS in Sect. 5.
Discussion and conclusion remarks follow in Sect. 6.

\section{The sample}
The study of  the continuum SED of 
 SSs in a large frequency range, from radio to X-ray,  
needs a large view of the data over all the spectral ranges.
More specifically, we assembled the sample data  by cross-checking the Belczy{\' n}ski et al. (2000) 
atlas with the ISO Data Archive (IDA),
and then by selecting those objects which showed the best-quality observations. 
Then, we have looked for complementary data in other wavelength regions in order 
to create an  extended   SED of the data for each object. 
This has allowed to better constrain several physical parameters. 
 At the same time, this has also  suggested to discuss 
some specific issues, e.g. the effect of intrinsic
flux variability on our modelling method, that we present at the end of this section.
Comparing the sample introduced in Table \ref{tab:lit} with  the original one studied  in Paper I, 
 notice that CH Cyg has been removed because
 we have  dedicated to it two  detailed papers  (Contini et al. 2009a,c). Moreover,
 a new object has been added: SS73 38. This SS has been chosen because  virtually ignored in literature,
and may thus represent a stimulating challenge for testing the  validity
 of our modelling approach.

\subsection{The ISO-SWS data}
The core of the data we exploit in this paper is composed of spectra recorded by the 
Short Wavelength Spectrograph (SWS - de Graauw et al. 1996) on board of the Infrared Space Observatory (ISO), 
which covered the wavelength range between 2.38 \mum\ and 45.2 \mum. 
In the present context, we employ them mainly as fine MIR continuum tracers. 
All these spectra have been presented and analysed for the first time in Paper I, the only 
exception being HM Sge (originally discussed 
by  Schild et al. 2001) and SS73 38 (still unpublished). Further technical details can be 
found in Paper I and references therein. The ISO-SWS journal of observations appears 
in Table \ref{tab:ISOjournal}.

\subsection{Complementary data from literature}
Some public surveys in the literature were particularly helpful in supplying observational data for the different 
objects of the sample. For the radio-mm range, the works by 
Purton et al. (1982), Ivison et al. (1992,1995) and Wendker (1995) have been fundamental. 

In the IR, despite the significant variability in the NIR range (mainly due to the Mira pulsations) 
that results in a non-perfect alignment at the edges of some photometric data points 
taken in different periods, we were able to build an unambiguous SED within the observation uncertainties. 
In this spectral range,  the 2MASS All-Sky Catalogue of Point Sources, 
the IRAS Catalogue of Point Sources, the MSX6C Infrared Point Source Catalogue (Egan et al. 2005), 
as well as the catalogue by Munari et al. (1992) have been fully exploited. The latter  provides photometric data for several SSs 
from the U (0.33 \mum) to the L (3.54 \mum) band. 

In the optical domain some papers presented photometric monitoring throughout the last years that allowed 
to choose data points in a few cases  very close in time to the ISO observations: 
in particular, we refer to Skopal et al. (1996, 2002, 2004, 2007), Arkhipova et al. (2004), 
Taranova \& Shenavrin (2000), Bergner et al. (1988), Munari et al. (1992), Gromadzki et al. (2009).

\begin{table*}
\centering \caption{ISO-SWS journal of observations. $\phi$ is the corresponding Mira pulsation phase, whenever ephemerids are known.\label{tab:ISOjournal}}
\begin{tabular}{c c c c c c c c c c c c c c }\\
\hline \hline 
Object & Coordinate center & ISO TDT & t$_{exp}$ & Observation date & $\phi_{Mira}$\\
& (J2000)& number & [s] & cal. / JD (2450) & -\\
\hline
BI Cru & 12:23:26.39 \, -62:38:16.5 & 25901615 & 1140 & 02-08-1996 / 297& 0.571\\
SS73 38 & 12:51:26.21 \, -64:59:58.8 & 60700908& 1912 & 15-07-1997 / 644& 0.288\\
V835 Cen & 14:14:08.99 \, -63:25:45.3 & 60702103 & 1140 & 15-07-1997/ -&-\\
H1-36 & 17:49:48.24 \, -37:01:27.0 & 32400609 & 1140 & 06-10-1996/ -&-\\
HM Sge & 19:41:57.11 \, 16:44:39.6 & 31901701 & 1140 & 01-10-1996/ 357&0.644\\
"& 19:41:57.06 \, 16:44:40.2 & 54700107 & 1912 & 16-05-1997/ 584&0.749\\ 
V1016 Cyg & 19:57:05.00 \, 39:49:36.0 & 35500977 & 1140 & 05-11-1996/ 398&0.214\\
"& 19:57:04.99 \, 39:49:35.9 & 55102706 & 1140 & 20-05-1997/ 588&0.612\\
"& 19:57:05.01 \, 39:49:36.4 & 74601883 & 1140 & 01-12-1997/ 783& 0.199\\
RR Tel & 20:04:18.52 \, -55:43:33.4 & 12402160 & 1062 & 20-03-1996/ 162& 0.664\\
"& 20:04:18.62 \, -55:43:33.1 & 54601206 & 1912 & 15-05-1997/ 583& 0.757\\
"& 20:04:18.49 \, -55:43:33.7 & 73402079 & 1140 & 19-11-1997/ 771& 0.245\\
V627 Cas & 22:57:42.07 \, 58:49:14.0 & 09604831 & 1044 & 21-02-1996/ -&-\\
R Aqr & 23:43:49.36 \, -15:17:40.3 & 18100530 & 1834 & 16-05-1996/ 219& 0.213\\
\hline \hline
\end{tabular}
\end{table*}

\subsection{Remarks about source variability}
\begin{figure}
\begin{center}
\includegraphics[width=0.40\textwidth]{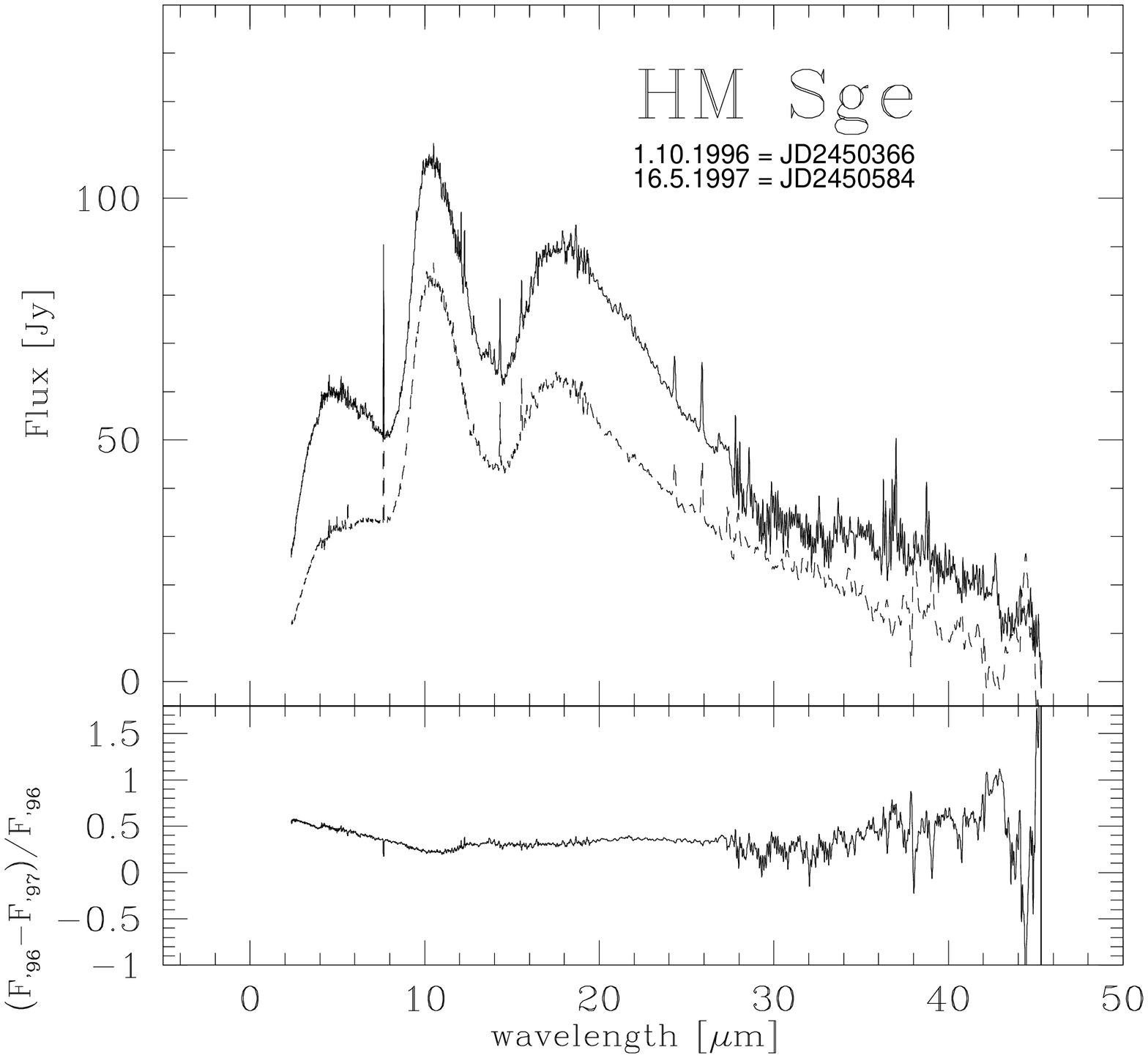}
\includegraphics[width=0.40\textwidth]{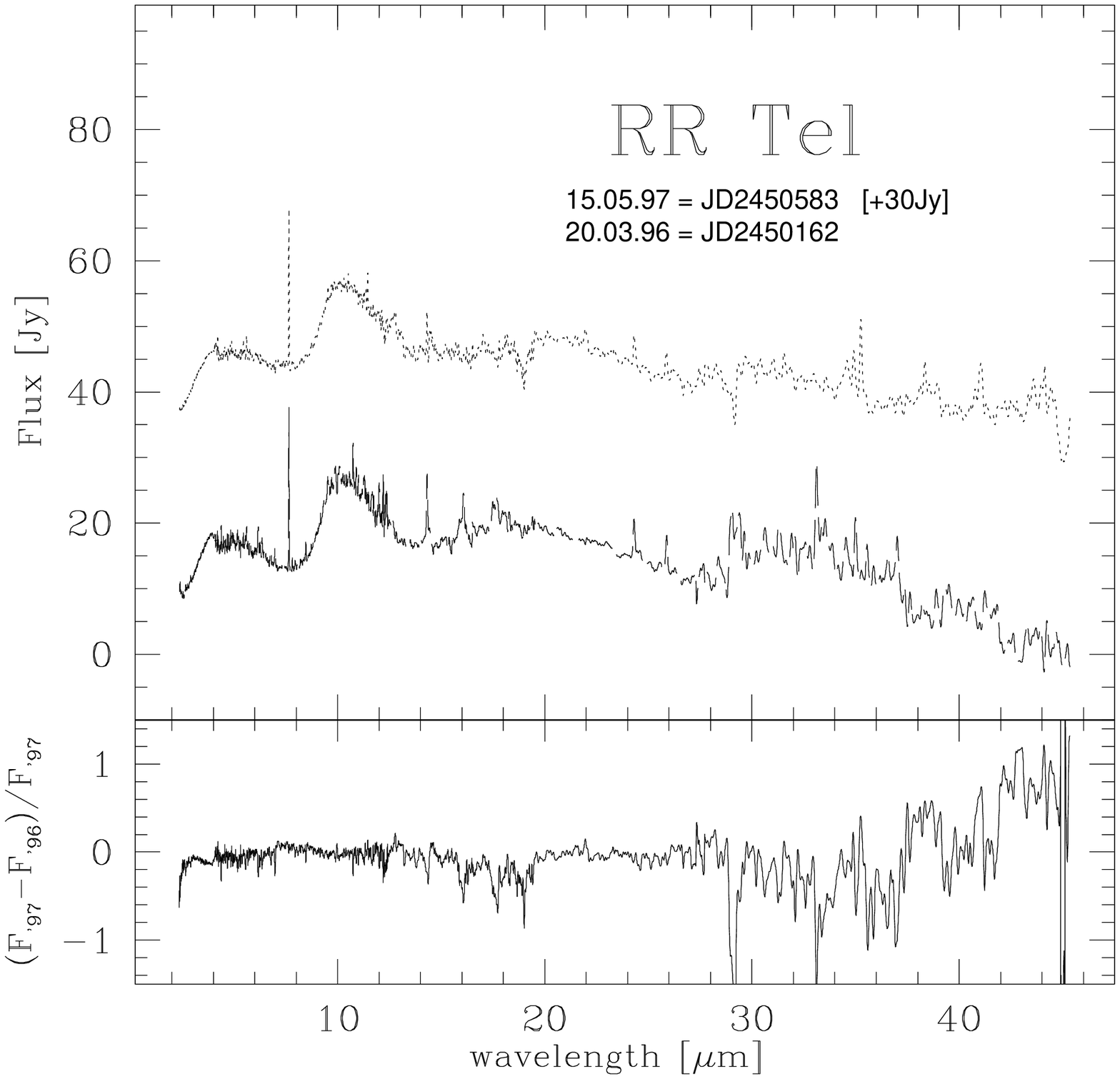}
\includegraphics[width=0.40\textwidth]{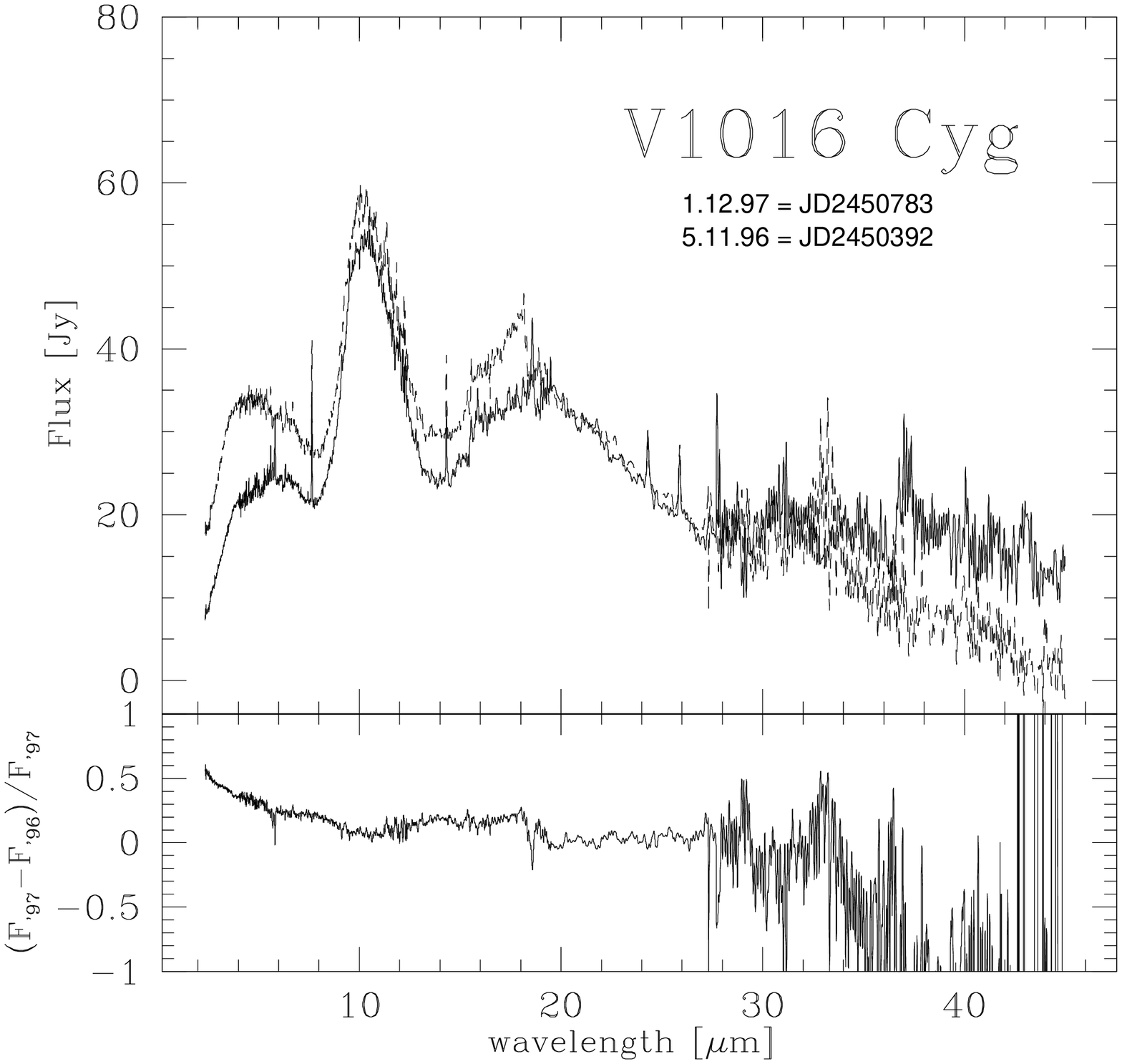}
\end{center}
\caption{Comparison of ISO-SWS spectra for three well-studied SSs. 
Top panels: plot of the ISO-SWS spectra taken in different date (1997: dashed line; 1996: 
solid line - note that for RR Tel the 1997 spectrum is arbitrarily shifted of 30 Jy to 
better distinguish the profiles). Bottom panels: relative difference of the spectra.\label{fig:var}}
\end{figure}

One of the most interesting aspect of SSs is the spectral and 
photometric variability. Nevertheless, it is highly unlikely that  
observations in such different ranges are coeval. 
So we try to understand whether, and for which purposes, the variability 
can be considered a second-order effect,  particularly for those objects known to have had outburst 
episodes, i.e. displaying the largest flux variations.

We  wonder whether the collection of data ranging from radio 
to UV, and belonging to a temporal interval that in some cases is as wide as a decade or more, might still 
be reliable to model  the SEDs. 

The first step has been to look at the literature, in order to understand how the community 
faces this fairly common problem. 

We have found that  several interesting results about radio and IR flux correlations were found 
out assembling observations 
taken within a temporal range of several years (e.g. Wright \& Allen 1978; Seaquist et al. 1993). 
For instance, the SEDs  of the jet emission from R Aqr and CH Cyg systems were built up by collecting 
data points taken in the years 1989-2002 (Galloway \& Sokoloski 2004, Fig. 4). 
Moreover, a study based on MIR observations of the SS CH Cyg (Biller et al. 2006) 
surprisingly found that data 
taken at very different epochs (e.g. IRAS PSC, IRAS LRS, ISO, photometric points from Taranova \& Shevranin 2000, 2004 
and Bogdanov \& Taranova 2001) are in good agreement with each other, despite CH Cyg being  known as a variable SS which 
has gone through several outburst episodes  even in the recent past.

The radio band deserves a special discussion: first, because the radio observations come from the '70 and '80 surveys, 
well before the ISO mission; second, because it is known that in SSs the radio emission can be significantly variable 
(Seaquist et al. 1984, 1993),  even more than in the IR. 
A study by Matthews \& Karovska (2006) of the near  system o Ceti (spatially resolved for the first time thanks 
to the Very Large Array), has shown that radio variability has not altered the flux densities by more than 30\% during 
the past 8.5 yr. 
Furthermore, an analysis of the continuum emission data at $mm$ and $cm$ wavelengths by Ivison et al. (1995), 
spanning a period of $\sim$ 4 years, has demonstrated that virtually all the objects of their large 
sample show flux density variations within a factor of 5. 
Translated in  SED diagrams (see $\S$4), this implies a vertical shift which still 
allows to discern between the main emission processes arising from the symbiotic nebul\ae{} (bremsstrahlung and synchrotron). 
An interesting   example of this is  given by CH Cyg, which in the late '80s showed bremsstrahlung and in the middle '90s 
synchrotron emission in the radio range (Contini  et al. 2009c).
Summarizing, we can confidently state, on the basis of current literature, that symbiotic variability does not modify the general properties of such 
systems as much as to invalidate our approach.

Whatever the case, we would like to verify this assumption directly, by fully exploiting the data for the 
unfortunately few, well studied objects  composing our sample. 
We have thus started verifying the agreement between the IRAS and ISO flux levels, although there is a time delay 
of $\sim$15 years. Moreover, whenever possible, we have also tried to join the IRAS and ISO spectra
to the NIR observations coming from long-lived photometric surveys: in some cases (as for 
HM Sge and V1016 Cyg) we have assembled a SED composed by data taken within a temporal range not larger 
than $\sim$10 days, from the U band to the upper edge ($\sim$45 \mum) of the ISO-SWS spectrum! 
Obviously, in this case the flux agreement is excellent.

Finally, some objects were observed by ISO-SWS more than once during its lifetime, 
making possible a direct comparison of the MIR spectral variability over different (tough unknown) Mira pulsation phases. 
In Fig. \ref{fig:var} we plot the ISO-SWS spectra for three well-studied SSs.
Despite some known instrumental effects, such as the sudden worsening of the spectrum quality longwards of  
$\sim$28 \mum \  due to the mediocre 
performance of the SWS band 3E at 27.5 - 29 \mum) we can recognize some interesting general trends. The strongest flux variations, 
whose relative ratio is  not larger than a factor of  0.6, arise from the cool star emission that dominates up 
to $\sim$ 5-6 \mum. 
The dust thermal emission (although not constant) varies within $\sim$ 0.3-0.4: this variability, as we discuss in  
Sect. 3, does not alter significantly our estimates of the dust shell sizes. For instance, by modelling the different ISO spectra for HM Sge, 
we find that the inferred dust 
shell size at $\sim$1000K varies by a factor $<$1.1, and the at $\sim$380K by a factor $<$1.2. 
Similarly, for V1016 Cyg, we find a variation of $\sim$1.1 for both the shells, well within the 
assumed uncertainty of our method.

\begin{figure}
\begin{center}
\includegraphics[width=0.5\textwidth]{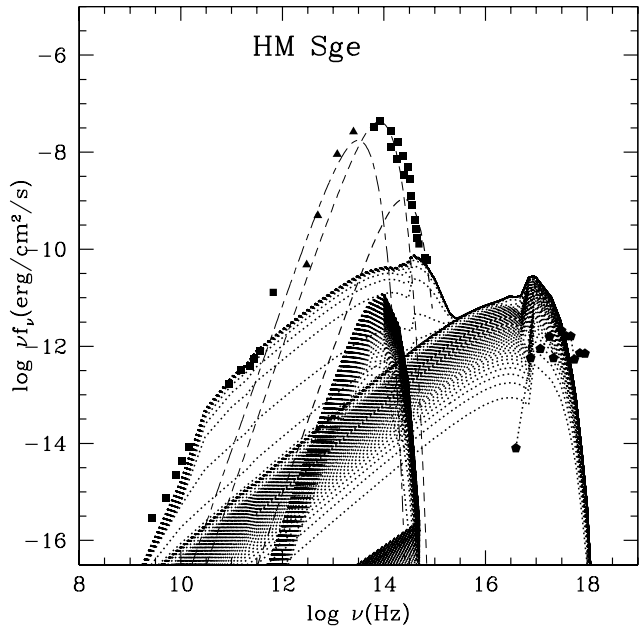}   
\end{center}
\caption{Dotted lines: bremsstrahlung and reprocessed radiation of grains
calculated consistently by SUMA in the different slabs
of the nebula downstream of the  reverse shock
in the HM Sge system. Dash-dotted lines: emission from the dust shells.
Dashed lines: emission from the red giant star.
\label{fig:slabs}}
\end{figure}

\begin{figure}
\begin{center}
\includegraphics[width=0.45\textwidth]{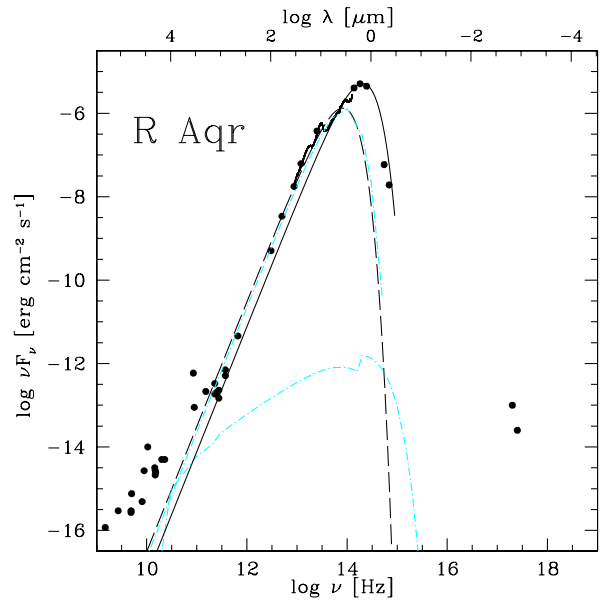}    
\end{center}
\caption{IR observational data vs. calculated SED of dust for R Aqr. Solid line: black body flux corresponding to 2800 K. 
Dashed line: black body flux corresponding to 1000 K. Dot-dashed cyan lines: the model.
Solid line: black body flux corresponding to 2800 K.
\label{fig:prov}}
\end{figure}

\section{The modelling}

In the past years, theoretical models as well as observations have shown that in SSs both the hot
and cool stars lose mass through strong stellar winds that collide
within and outside the system, hence creating a complex
network of wakes and shock fronts that  result in a complicated
structure of gas and dust nebulae (Nussbaumer 2000).
This  shock network  results from the 
interaction of the fast WD wind with the slow, dense outflow from the Mira, and corresponds to 
the  colliding wind configuration of  
Girard \& Willson (1987), which is  a theoretical basis to the more realistic
picture of  shock fronts disrupted by hydrodynamic instabilities. 

In the light of this scenario, one can consider that
two shock-fronts develop from the  collision of the winds between 
the stars (Girard \& Willson 1987, Kenny \& Taylor 2005): one strong shock-front facing the WD 
(the reverse shock) and the other weak one facing the red giant.
Similarly, in the extended circumbinary region, two shock fronts develop
from the head-on-back collision of the winds: one expanding outwards
and the second, virtually negligible, facing the system center (see the sketch presented by Contini et al. 2009c, Fig. 2). 
In the following, we refer to the two main shock fronts: namely, the reverse 
between the stars and the  expanding  outwards the system.
The expanding shock front is disrupted by  collision with the ISM inhomogeneities.
 
The gas entering the  shock front is thermalized up to a maximum temperature of
T=1.5 10$^5$ (\Vs/100 \kms)$^2$  (where \Vs\ is the shock velocity)
in the  region  immediately behind the
discontinuity. 
A high temperature  zone appears in the downstream region.
Then, the gas recombines following a  relatively high cooling rate due to the
high pre-shock densities characteristic of symbiotic nebulae and to compression downstream.

In the calculation code SUMA, the downstream region is cut in many plane-parallel slabs in order to follow as smoothly as possible
the trend of the physical conditions, particularly, the temperature.

The calculations of the fluxes  in the different slabs of the gas downstream
is shown in Fig. 2 in the case of HM Sge, starting by the slabs at higher temperatures.
Shock velocities
 can reach $>$ 400 \kms in the region between the stars by head-on
collision of the winds (Paper I). The high temperatures downstream lead to emission in the high frequency domain
up to X-rays.  
On the other hand, the radiation from the WD, even at outburst,  heats the gas   to no more than 1-3 10$^4$ K
in a relatively large zone downstream,   leading
to the  bump in the optical range and to the high bremsstrahlung for 10$^{10}$ $<$ $\nu$ $<$ 10$^{15}$ Hz.

The input parameters of the code are those relative to the shock: 
the pre-shock density, \textit{\n0},
the shock velocity, \textit{\Vs}, the pre-shock magnetic field, \textit{\B0}, 
and those relative to photoionization: the hot star ionizing radiation flux and its 
colour temperature \textit{\Ts}.
The   abundances  of He, C, N, O, Ne, Mg, Si, S, Ar and Fe, 
relative to H, the dust-to-gas ratio $d/g$ and the geometrical thickness of the nebulae $D$, 
are also accounted for.  \\
The key parameter is obviously the shock velocity, \Vs. For high-velocity shocks, the very 
high temperature reached in the post-shock region leads to the X-ray emission observed 
in several SSs. Moreover, broad  strong lines can also  be observed along the whole 
electromagnetic spectrum, particularly the coronal lines in the 
infrared  (Paper I): therefore, from the spectral point of view, 
different line profiles trace different velocity regimes, 
allowing to highlight the different physical conditions within a SS.

The dust grains  are
heated both radiatively by the primary and secondary radiation flux
from the hot star and from hot gas  downstream, respectively,
and collisionally by the gas in each slab  downstream.
 In  a shock-dominated regime,
collisional processes dominate and the grains can reach relatively high
temperatures. The intensity of the dust re-radiation emission peak depends strongly
on the dust-to-gas ratio, while the frequency corresponding to the peak
depends on the shock velocity.

Furthermore, since the matter is highly inhomogeneous at the shock fronts because of 
instabilities at the fluid interface (e.g. the Rayleigh-Taylor, Kelvin-Helmholtz, Meshkov-Richtmyer), 
different physical conditions should be accounted for in each one of the symbiotic system, 
particularly regarding the density (e.g. in BI Cru - Contini et al. 2009c).

\section{The continuum SED}

In this paper we  deal  with the continuum SED of a sample of D-type SSs.
The data in the X-ray, UV, optical, IR, and radio ranges are modelled by the
free-bound and  free- free flux (hereafter bremsstrahlung) calculated consistently
by each model. 

The IR continuum SED in SSs give some direct information about the
system components, in particular, of the dust shells. 
Moreover, in  the IR range, reprocessed radiation by dust grains coexisting with the gas
in the nebulae is accounted for.
It was found by modelling  previous SSs
that the peak of dust re-radiation calculated consistently with bremsstrahlung
seldom  emerges  over it, adopting even high $d/g$ ratios.
The  SEDs  in the IR are explained
by the summed fluxes from dust at different temperatures, which are
 well recognizable,  indicating  that they most likely  correspond to different
shells.

In the radio and in the X-ray ranges, the continuum  accounts for the emission from the shocked nebulae.
 The  WD flux appears in some objects
in the U band and in the UV range, even if the shocked nebulae also contribute (see Contini et al. 2009c
for  detailed modelling).

\begin{table*}
\centering \caption{the references of the data in Fig. \ref{fig:seds}\label{tab:ref}}
\small{
\begin{tabular}{ccccccccccccc}\\
\hline
\hline
  symb & radio - mm   & IR   & optical - UV \\
\hline
    BI Cru & Ivison et al. 1995 & IRAS; ISO; 2MASS & Rossi et al. 1988; Gromadzki et al. 2009 \\
    SS73 38 & Wendker et al. 1995 & IRAS; ISO; 2MASS & Gromadzki et al. 2009\\
    V835 Cen &Purton 1982; Ivison et al. 1995 & IRAS; ISO; Feast et al. 1983& - \\
    H1-36 &Purton 1977; Ivison et al. 1995 & IRAS; ISO&Allen 1983 \\
    HM Sge & Purton 1982; Ivison et al. 1992 & IRAS; ISO; Taranova et al. 2000 & Arkhipova et al. 2004 \\
    V1016 Cyg  &  Purton 1982; Ivison et al. 1992& IRAS; ISO; Taranova et al. 2000 & Taranova et al. 2000 \\
    RR Tel  & Purton 1982; Ivison et al. 1995&IRAS; ISO; 2MASS & Kharchenko 2001; Gromadzki et al. 2009 \\
    V627 Cas  & - & IRAS; ISO & Bergner et al. 1988\\
    R Aqr  &  Purton 1982; Ivison et al. 1992 & IRAS; ISO; 2MASS & Kharchenko 2001; Gromadzki et al. 2009\\       
\hline
\end{tabular}
}
\end{table*}

\subsection{The continuum from gaseous nebulae}

The  continuum SED of SSs gives more direct informations about the
system components than the line spectra do. However the latter are more constraining
the models.  For instance, the relative chemical abundances
for each model can be found out modelling the line ratios, while only dramatic changes in the abundances 
of carbon, nitrogen, and oxygen, 
that are strong coolants, can  strongly alter the continuum SED.

Since the observations are taken  at the Earth, while the models are calculated 
at the nebula, to compare the models with the data,
we define the factor $\eta$  =({\it ff} r/d)$^2$, where r is the  distance of the nebula
from the SS center, d the distance to Earth,
and {\it ff}  ~the filling factor.
The $\eta$ factors, depending on  the distance of the nebulae from the system center,
not only  constrain the models but also provide precious informations about
the  radius of  shocked nebulae and dust shells within the system.

The results of modelling for single objects are presented in the next section.
For some objects of the sample (i.e. RR Tel, HM Sge, R Aqr, H1-36, and BI Cru)
we already analyzed in detail the emission line spectra in previous papers.
Therefore, for the sake of consistency and also to verify the validity of our results about the 
 continuum, we use the models consistently calculated on the basis of the emission line ratios. 
We are also aware that in the last years new data have appeared; moreover, the SUMA code
has been updated. 
Therefore, we have also run new models based on those previously adopted which take into account the new
features.
For the other objects which  either have been less observed or  for which we have not considered 
explicitly the emission line spectra (V627Cas, V835Cen, SS38 and V1016 Cyg) we have used 
as a basis the main models which explained the SED of other objects, adopting always one  model 
representing the shocked nebula downstream of the reverse
shock between the stars, and the other representing the expanding shock.

\subsection{The dust shells}

Dust shells with  different temperatures were discovered
by the detailed modelling of single  D-type SSs 
(e.g. Paper I, Angeloni et al. 2007b, Contini et al. 2009b)

According to hydrodynamic model atmospheres, the  Mira variable pulsations
lift up matter from the stellar surface
and extend it  to  the atmosphere, which is modified by the shock waves
created by stellar pulsations.
Locally, the passing shocks cause a temporary variation of the thermodynamical
conditions, influencing the formation of molecules and dust grains (Hoefner 2009).
In fact the shock waves  lead to a levitation of the
stellar atmosphere, providing a cool and dense environment as required for an
efficient formation and growth of dust grains.

The newly formed  grains are accelerated
by the stellar radiation field and initiate a slow massive outflow by transferring
momentum to the surrounding gas.
More particularly, while
the interaction of a particle with a gravitational field depends solely on the particle mass,
its interaction with  a radiation field depends on its composition, structure, size, and density,
as well as on the radiation wavelength.
Thus, if in the outer layers of a star there are  particles
that  are exceptionally  absorbent at the leading wavelength of the photon -as determined by the
temperature - then for these particles the radiation pressure might just overcome gravity (Prialnik 2000
).
The result would be an outer acceleration leading to a mass outflow of such particles and other
entrained by them.

These particles  are accelerated by the radiation pressure. Consequently
a  dust  shell is  ejected from the star.
The dust  shells can then be destroyed by evaporation, when
the grains reach temperatures
$\geq$ 1200 K,  or by  sputtering of the grains throughout strong shocks.

In addition IR observations have shown the existence of a cool emission
component which
indicates dusty circumstellar material.

We would like to investigate in detail the dust-to-gas ratio by mass in the shells.
This is an interesting issue, not only in the specific case of symbiotic systems,
but in the understanding of IR emission from starburst galaxies and luminous IR galaxies
(e. g. Contini \& Contini 2007), which is generally attributed to supernovae (e. g. Dweck  1987).

 Moreover, extinction increases with  dust mass and dust is created in the atmosphere
of the giant star. On the other hand, the grains cannot survive close to the WD because
they evaporate at temperatures $\geq$ 1500 K. So extinction of the Mira is generally greater than that
of the WD in several SSs (one can notice it in H1-36 and HM Sge also looking at our Fig. 4, 
where the cool star is virtually unnecessary in modelling the data around log$\nu\sim$14.5.). 
However, in a system observed edge-on the WD is periodically occulted by
the red giant star and by the network of shocked nebulae created by collision of the winds.
This leads to complex light curves as those observed from e.g. CH Cyg.

Previous models for circumstellar shells deals with isolated stars. 
For instance Young et al. (1993) 
 adopt  a simple model of shell evolution, involving the interaction
of the expelled material with the ISM. For oxygen rich Mirae,
the models of Le Sidaner \& Le Bertre (1996)  consider a dust shell surrounding  
a central star
which radiates as a black body. All the models lead to  dust temperatures $\geq$ 800 K.

Our model is  similar to that used for the shells in the CH Cyg system (Contini et al. 2009c),
namely, the dust+gas shell moves outwards the Mira and a shock front forms on the
external edge. The radiation flux from the hot star
 reaches the shell on the shock front.
 We proceed as for the nebulae,
considering many downstream slabs and calculating the physical conditions in each of them.
Radiation  is calculated by radiation transfer
throughout the different slabs. 
 
We adopt a shock velocity of $\sim$ 30 \kms.  The shocks form  by head-on collision 
of the shell with the wind from the WD between the stars and  by head-on-back collision
with the  giant wind or by collision with the interstellar matter
beyond the binary system. The code SUMA is used for the calculations.
The other significant input parameters used for the calculations are \n0=6 10$^9$ \cm3, 
\B0=3 10$^{-3}$ gauss, similar to those adopted by  Contini et al. (2009c, Table 1).

The results show that dust grains are heated to $\sim$ 1000 K collisionally by the   
gas   which reaches  temperature T$\sim$ 10$^4$ K
throughout the shock front and downstream.
Recall that shells with  dust  temperatures of $\sim$ 1000 K are present in late-type
stars (Danchi et al. 1994) as well as in SSs, showing
that collisional heating  dominates.
The   radiation from the WD  will  most easily destroy the grains by heating them beyond
the evaporation critical temperature ($\sim$ 1500 K).

The best fit  to R Aqr continuum SED in the IR (Fig. 3) is obtained by  $d/g$ 
by mass $\geq$ 0.1, 
which is of the order of the dust-to-gas ratios found in SN and  by a factor 
of $\sim$ 10$^3$ higher than in the ISM.

The  reprocessed emission by dust is well reproduced by the black body (bb) flux
at 1000 K because only  the first few slabs of the shell close to the shock front, 
emit  a strong flux.
In the following slabs, the  grain temperature
rapidly decreases following the  high cooling rate of the  gas.
So, generally,  the emission from  a shell
 is well approximated by a bb  flux at a certain temperature.
In the following, we represent the shell  emissions by  bb fluxes
at the temperature best fitting the data in the continuum SED.

Beyond the Mira, on the side opposite to the WD, the shock accompanying the shell
 accelerates throughout the decreasing density slope in the Mira extended  atmosphere.
 At  \Vs=100 \kms and \n0=2 10$^7$ \cm3 the grains are collisionally heated to $\sim$ 440 K,
as observed in different objects.
\begin{table*}
\centering \caption{The models adopted for the single objects  described in Sect. 5 \label{tab:suma}}
\begin{footnotesize}
\begin{tabular}{lllllllll}\\ \hline  \hline\\
\ \textbf{Object} & \Vs & \n0 & \B0 & \Ts & U & $D$ & $d/g$\\
\hline
\ \textbf{BI Cru} &&&&&&& \\  
\ rev        & 400 & 1.e5 & 1.e-3 & 2.65e4 & 1 & 1.e14 & 4.e-4\\
\  exp       & 150& 1.6e3& 1.e-3& 2.65e4& 25& 1.e15& 4.e-4\\
\ lobes      &210& 2.e4& 1.e-3& -&-& 1.e18&4.e-4\\
\ \textbf{SS73 38} &&&&&&&\\
\ rev        & 400 & 1.e5 & 1.e-3 & 2.65e4 & 1 & 1.e14 & 4.e-4\\
\ exp    & 70 & 3.e4 & 1.e-4 & -   & -  & 5.e15&4.e-5\\             
\ \textbf{V835 Cen} & &&&&&&  \\
\ rev        & 400 & 1.e5 & 1.e-3 & 2.65e4 & 1 & 1.e14 & 4.e-4\\
\ exp    & 70 & 3.e4 & 1.e-4 & -   & -  & 5.e15 & 4.e-5\\             
\ \textbf{H1-36}  &&&&&&&&\\
\ rev        & 140 & 2.5e5 & 1.e-3 & 1.5e5 & 2.e-3 & 2.8e12 & 4.e-4\\
\ exp        & 70  & 3.5e3 & 1.e-3 & 1.5e5 & 2.5e-3 & 1.e15 & 4.e-4\\
\ \textbf{HM Sge} &   &&&&&\\
\ rev  & 500 & 5.e5& 1e-3 &1.6e5& 1.& 1.4e14&4.e-5\\
\ exp      &210& 2.e4& 1.e-3& -&-& 1.e18&4.e-4\\
\ \textbf{V1016 Cyg}  &&&&&&&&  \\
\ rev   & 500 & 5.e5& 1e-3 &1.6e5& 1.& 1.4e14&4.e-5\\
\ exp    & 70 & 3.e4 & 1.e-4 & -   & -  & 5.e15 & 4.e-5\\             
\ \textbf{RR Tel} &&&&&&\\ 
\ rev1& 80 & 1.e6 & 1.e-3 & 1.4e5 & 0.2 & 5.5e13 & 4.e-4\\
\ exp&  45 & 2.e6 & 1.e-3 & 1.4e5 & 2.e-4 & 1.e14 & 4.e-4\\
\  rev$_{XX}$     & 500& 1.e6&1.e-3 &1.4e5 & 0.2& $<$ 5.9e13& 2.e-4\\
\ \textbf{V627 Cas} &   &&&&&&\\
\ rev        & 400 & 1.e5 & 1.e-3 & 2.65e4 & 1 & 1.e14 & 4.e-4\\
\ exp    & 70 & 3.e4 & 1.e-4 & -   & -  & 5.e15 & 4e-5\\             
\ \textbf{R Aqr}  &&&&&  &  \\
\ rev$_{HII}$ & 120 & 6e4 & 2e-3 & 8e4 & 6e-3 & 2e14 & 4e-4\\
\ exp$_{150}$ & 150 & 4e3 & 1.e-4   & -& -& 2.5e15 & 4e-4\\
\ exp$_{300}$& 300 &1.e3 & 1.e-4& &-&-& 4.e-4\\
\hline
\end{tabular}
\end{footnotesize}
\end{table*}

\section{Results}


\begin{figure*}
\begin{center}
\includegraphics[width=0.32\textwidth]{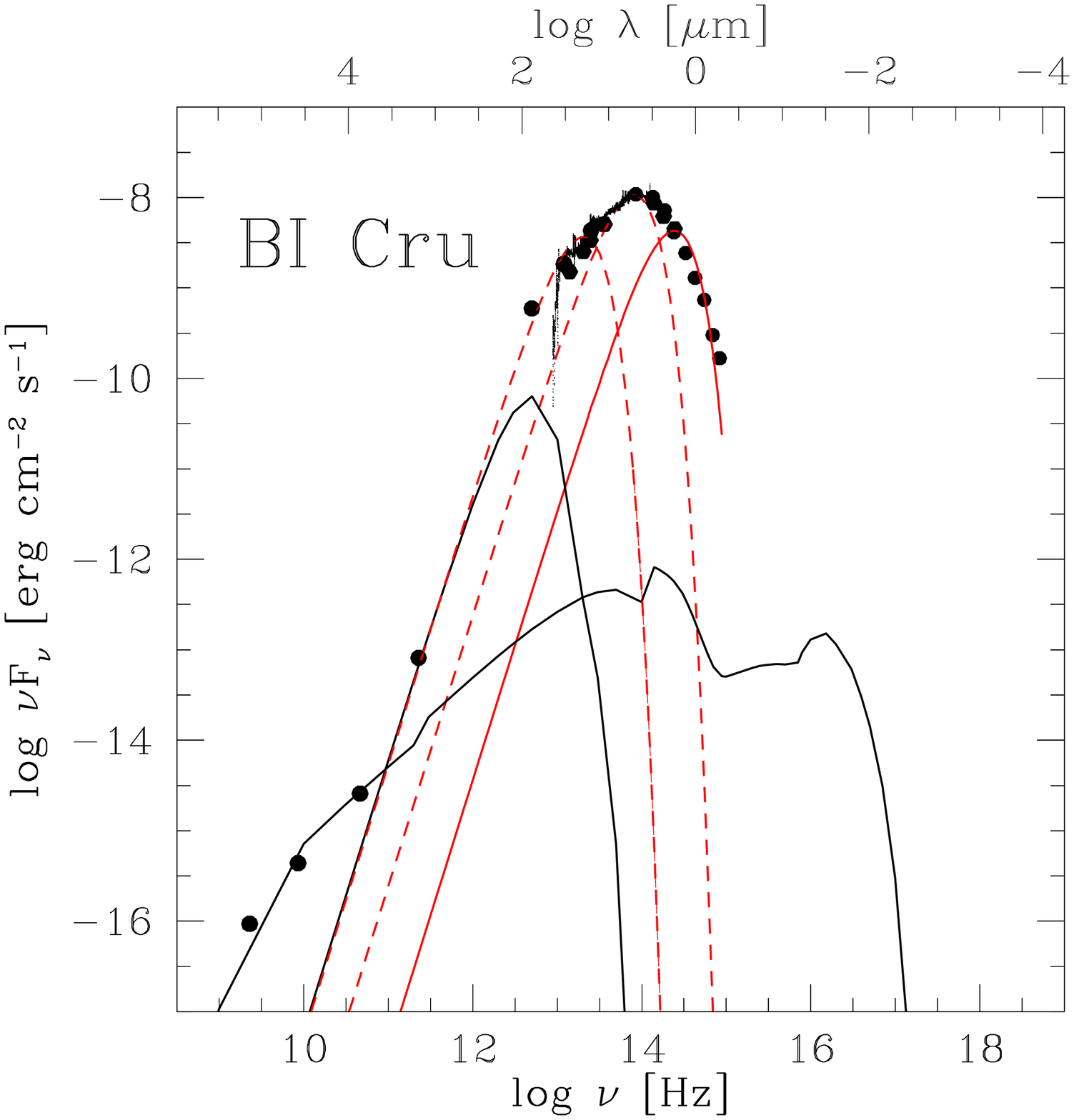}  
\includegraphics[width=0.32\textwidth]{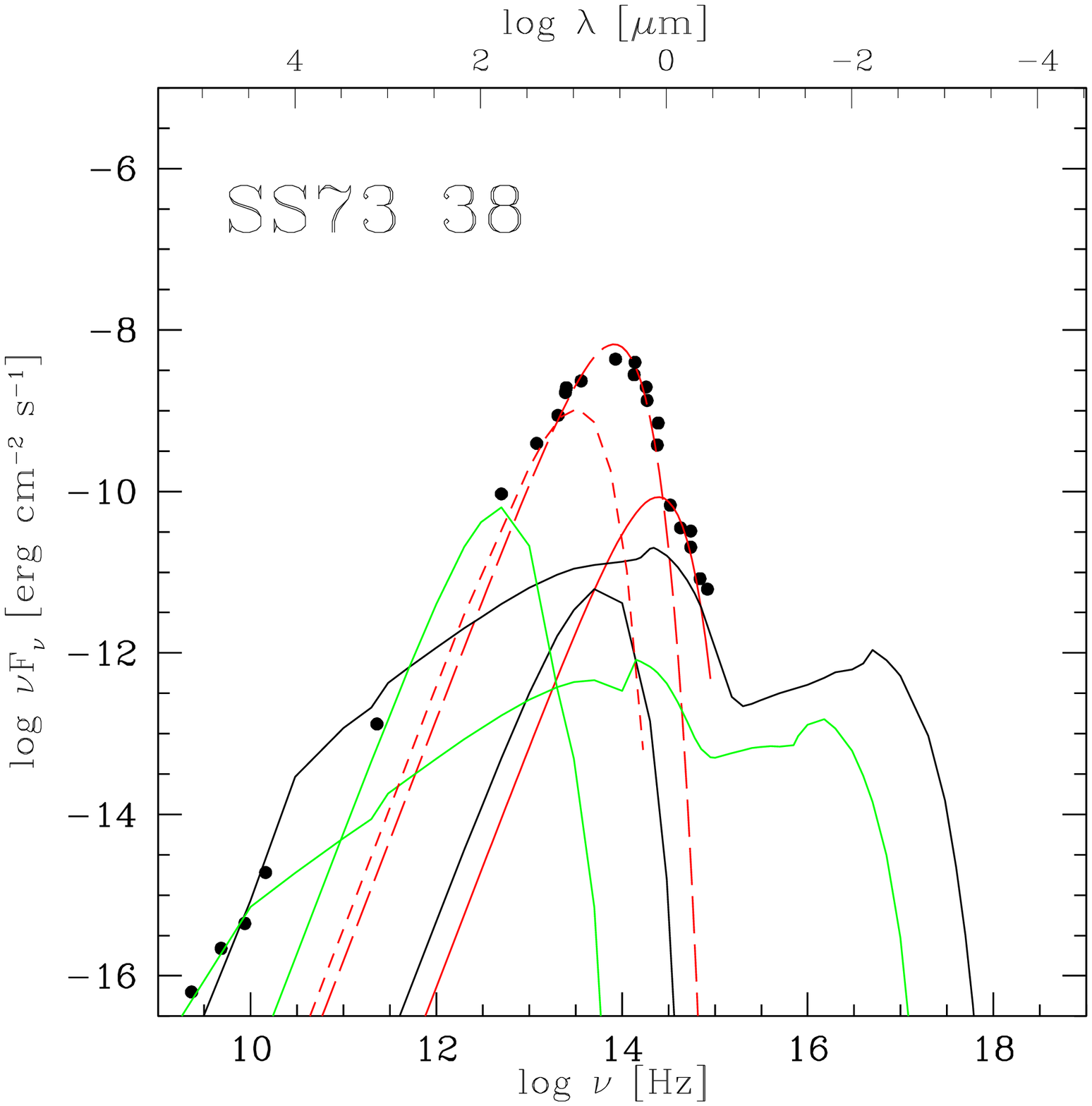} 
\includegraphics[width=0.32\textwidth]{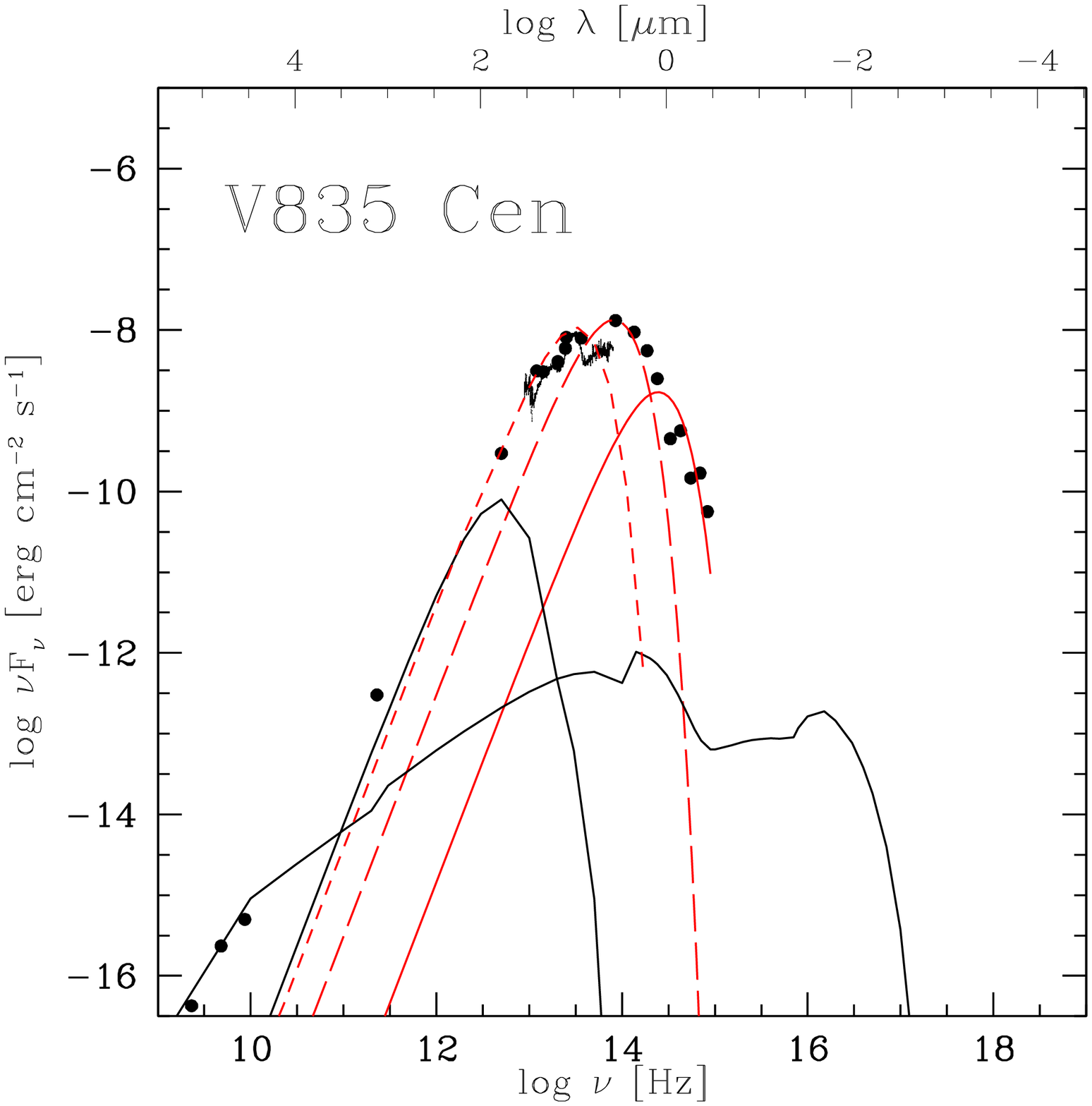}  
\includegraphics[width=0.32\textwidth]{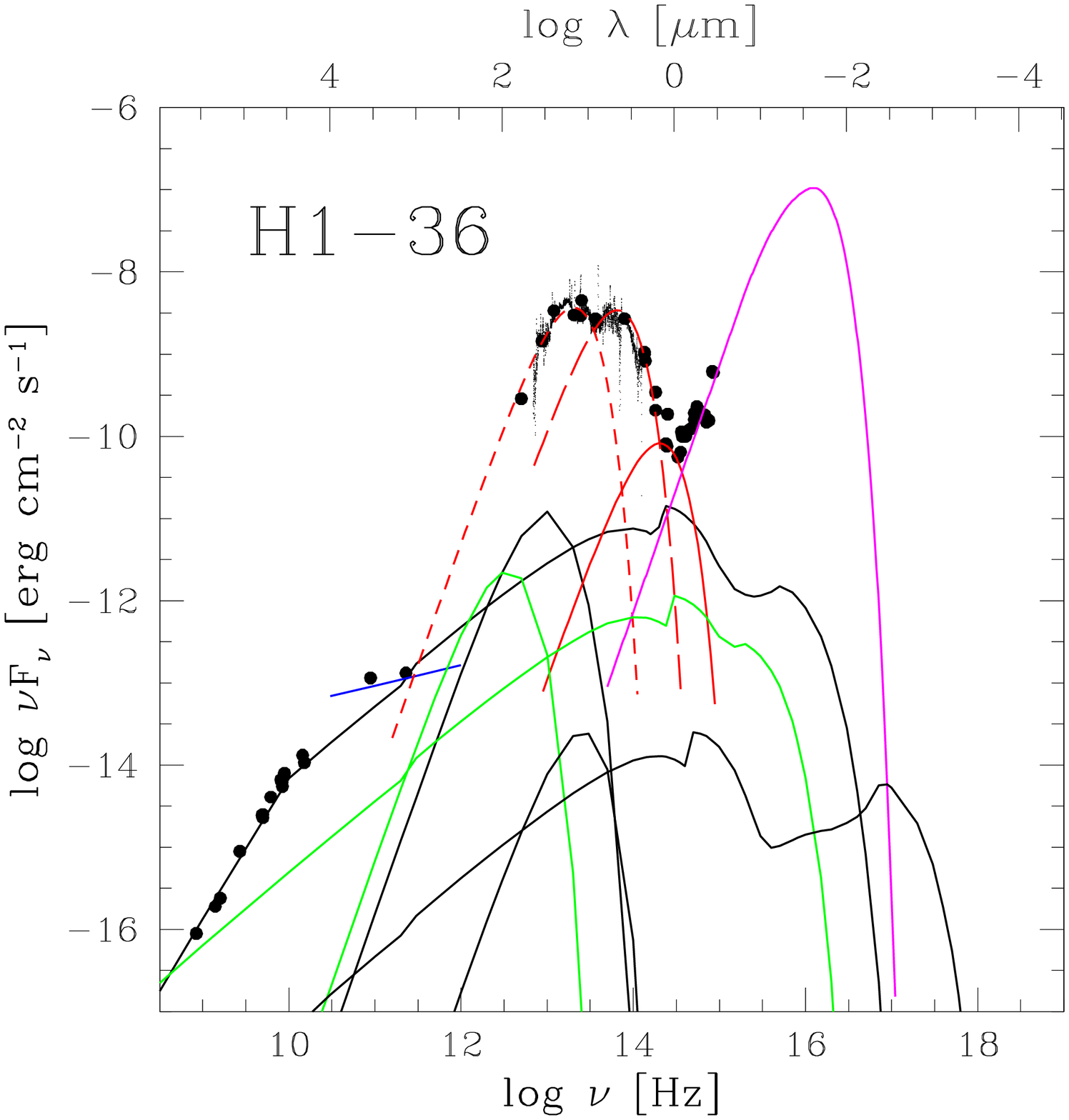}  
\includegraphics[width=0.32\textwidth]{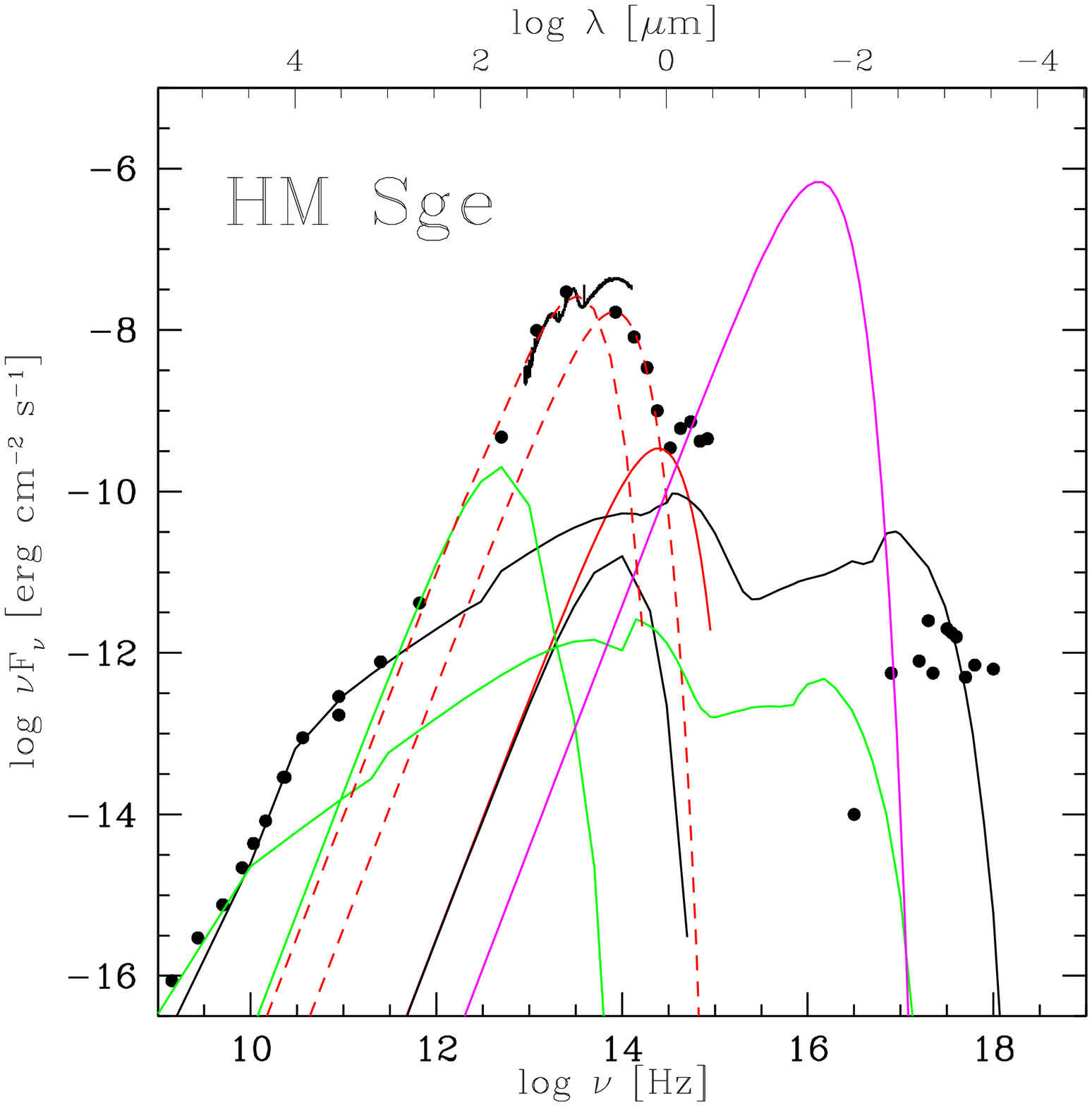}  
\includegraphics[width=0.32\textwidth]{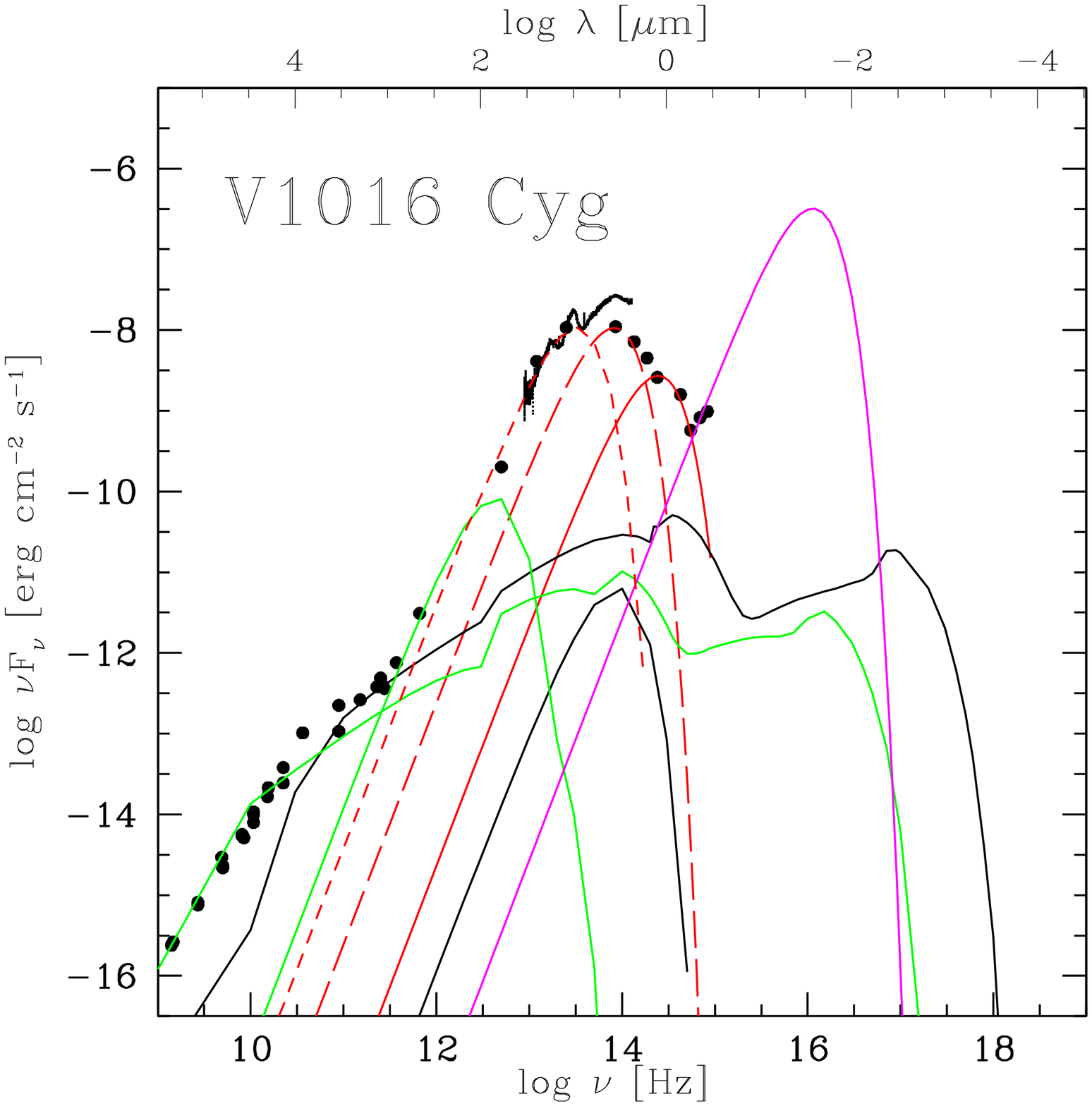}  
\includegraphics[width=0.32\textwidth]{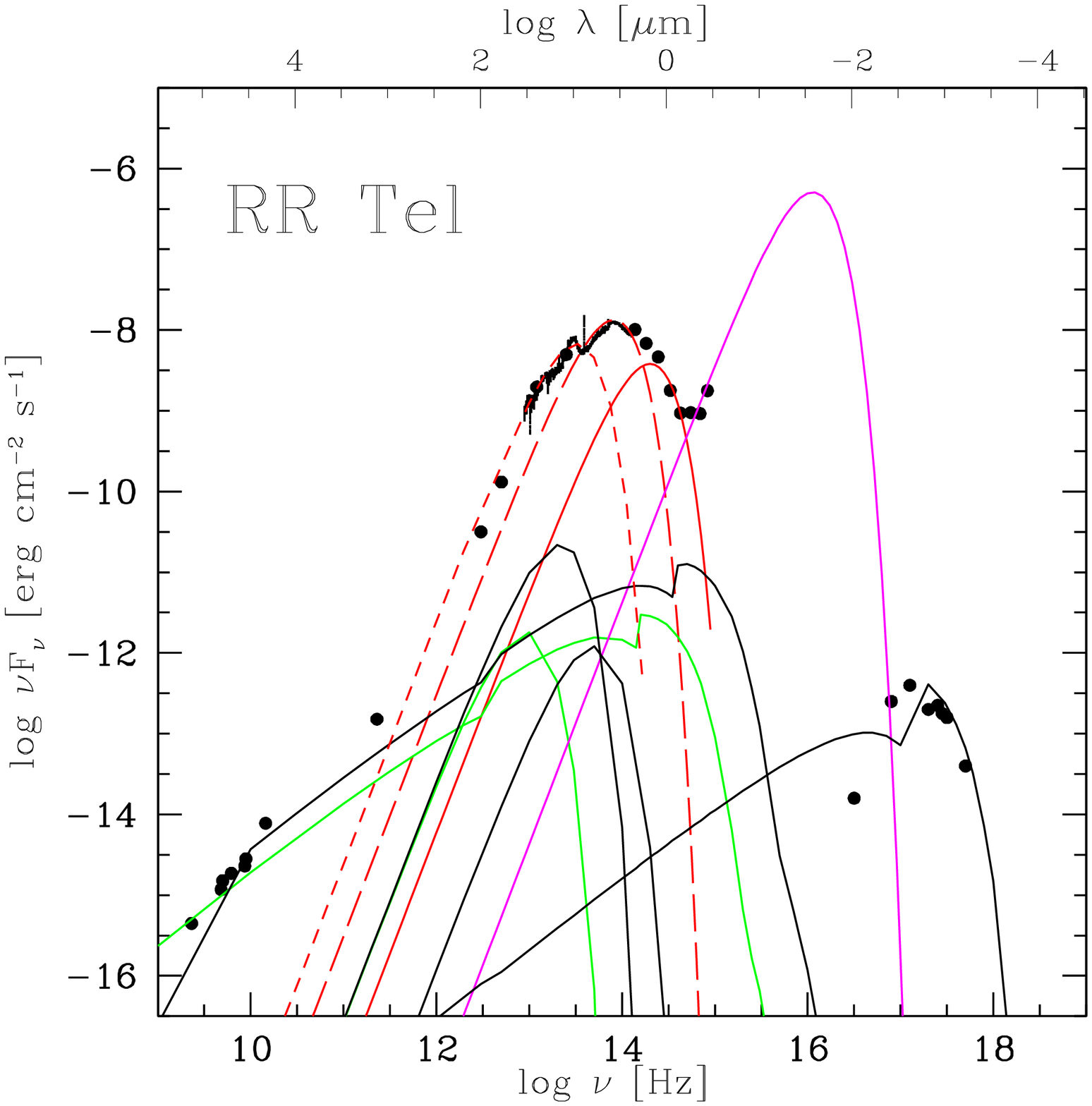}  
\includegraphics[width=0.32\textwidth]{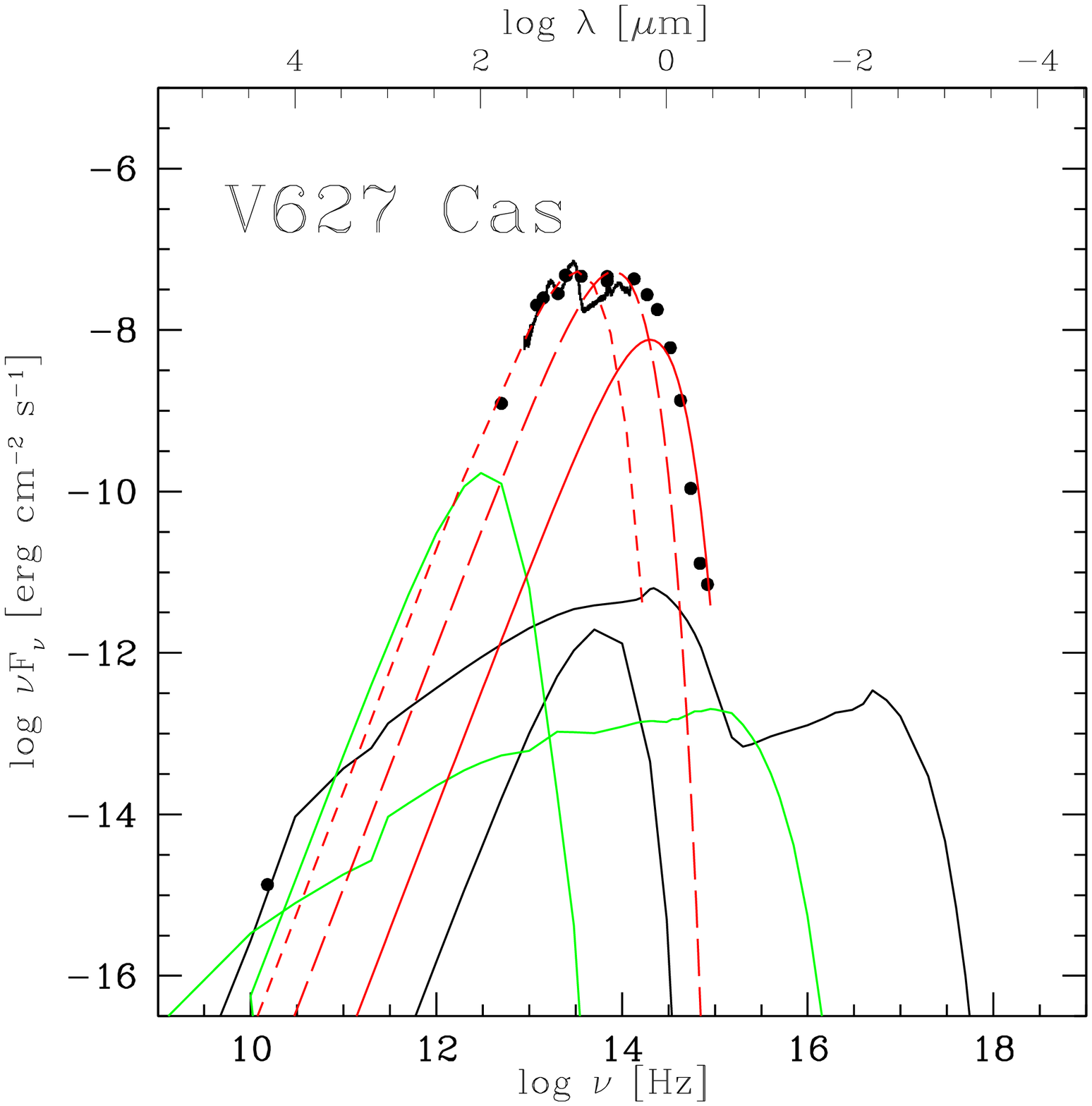}  
\includegraphics[width=0.32\textwidth]{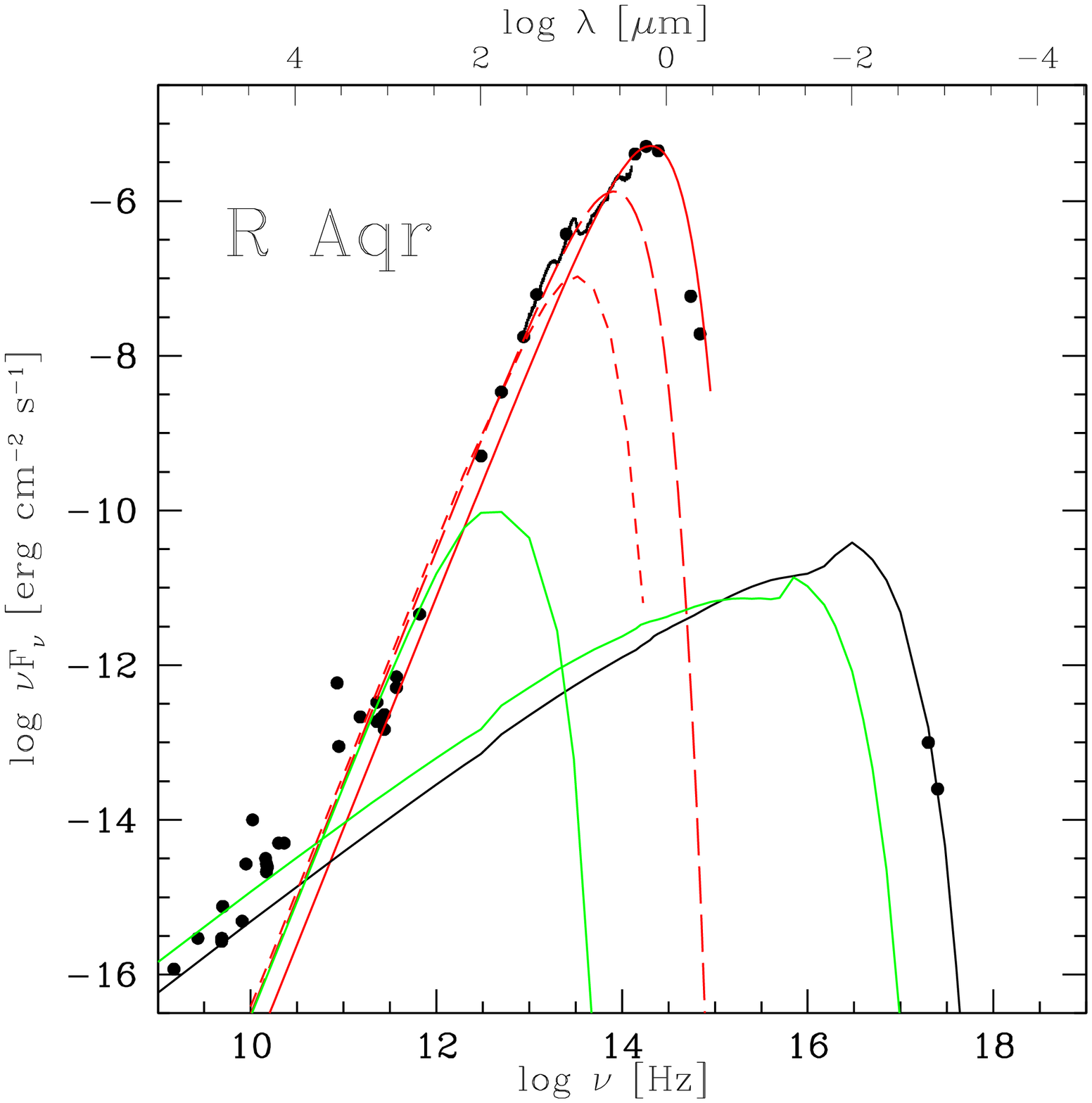}  
\end{center}
\caption{The modeling of single objects, from radio to X-ray range. 
Black dots: the data;
Red: solid: bb from the red giant;
long-dash: bb from the hot dust shell;
short-dash: bb from the cool dust shell;
Magenta: bb from the WD;
Black: bremsstrahlung + dust re-radiation downstream of the reverse shock front;
Green: bremsstrahlung + dust re-radiation downstream of the  expanding shock front;
\label{fig:seds}}
\end{figure*}

We present in Fig. \ref{fig:seds} the results of modelling the continuum SEDs of D-type SSs.
The input parameters adopted  for the best fit of models to  data (Table 4)  are obtained  by
cross-checking the line spectra in most of the objects. They
represent  the physical conditions in the different  nebulae within the SS.\\
Some significant results appear in Table \ref{tab:res}
In column  2 the best fit  temperature of the Mira is given, in columns 3 and 4 the radius of the
reverse and expanding shocks appear,  followed in columns 5 and 6 by the radius of the dust shells
at 1000K and 400 K, respectively.

A first look at the diagrams in Fig.  \ref{fig:seds} reveals that  the component types
of SSs are similar. A reverse and an expanding shocks are usually present, as well as multiple dust shells. 
Virtually in all the objects, the peak of the dust shell at 1000 K  shows the maximum flux 
throughout the entire continuum SED, from radio to X-ray. 

The self-absorbed free-free radiation flux in the radio range corresponds to the reverse shock. 
On the other hand, the expanding shock is responsible for  the radio emission in  the low
frequency tail.  Moreover, between log($\nu$)=16.6 and 18 the X-rays  can be strongly absorbed
(Zombeck 1990).\\

Brief  notes on the single objects are presented in the following.

\centerline{\bf BI Cru}

A careful analysis of BI Cru as a colliding-wind SSs has been recently described by Contini et al. (2009b). 
We refer to that  paper for any further details about models and results. 
Here it is worth mentioning that the IR SEDs show the presence of two, relatively cooler, dust shells at 800 K and 250 K.
BI Cru has a very unusual near-IR spectrum, with  strong CO in emission
(Whitelock et al. 1983, Fig. 3)
 which suggests that something other
than dust may  give a significant contribution to  the flux. 
For instance the presence of SiO  molecules in  D-type SSs (Paper I) was predicted from the intensity
ratio of characteristic Si lines. Unfortunately, no millimeter spectra are available for this object.\\

\centerline{\bf SS73 38}

The data in the radio range are well reproduced by the same models
adopted for HM Sge. The datum at the 
highest frequencies  represents the summed contribution of the flux
from the red giant (T=3000 K) and bremsstrahlung from the reverse-shock nebula.

Also in this case, two dust shells are found. Nonetheless, the dust temperatures are slightly higher than in BI Cru, 
being respectively at 1000K and 400K. These appear to be the standard values for dust shell 
in SSs as revealed on the basis of our analysis.\\

\centerline{\bf V835Cen}

Only the model with \Vs=210 \kms appears in the modelling.
The spectra  from V835 Cen by Gutierrez-Moreno \& Moreno (1996) show the presence of the
[OII]3728 line which indicates relatively low densities. In fact, the
critical density for collisional deexcitation is low ($\sim$ 3000 \cm3).
Moreover, the presence of the [FeVII] lines with a [FeVII]6087/[OIII]5007
$\sim$ 0.07 is compatible with shock velocities \Vs $\sim$ 200 \kms
(see the grid calculated  for the shock dominated  spectra by Contini \& Viegas 2001).
This leads to a model similar to that  adopted for the lobes of BI Cru.
The model is shock-dominated, i.e.  not constrained by the
WD temperature.
The red giant flux is reproduced by a bb corresponding to T=3000 K.
Two dust shells are revealed, at 1000 K and 400 K respectively.\\

\centerline{\bf H1-36}

This SS was modelled in detail by Angeloni et al. (2007b).
We refer to that  paper for any further details about models and specific results.
The Mira is at T=2500 K and the dust  shells are at T=850 K and at 250 K.\\

\centerline{\bf HM Sge}

The reverse model is described in Paper I.
The expanding model, adopted phenomenologically, is similar to that used
for BI Cru lobes.
 The Mira temperature is approximated by T= 3000 K
and also in this case the dust shells are at T= 1000 K and 400 K.
The X-ray data (Kwok \& Lehay 1984) indicate that there is a strong
absorption in the X-ray domain.
Within the 10$^{16}$ -10$^{18}$ Hz range X-ray are heavily absorbed 
with absorption cross sections increasing towards lower frequencies (Zombeck 1990),
thus improving the fit  to this object, as shown in Fig. \ref{fig:seds}. \\

\centerline{\bf V1016 Cyg}

 The line spectrum presented by Schmid \& Shield (1990) shows lines from different 
ionization levels and from different elements. The line ratios are  very similar
to those  of the HM Sge line spectrum, therefore we use
the models adopted for HM Sge (Paper I).
The temperature of the Mira is approximated by T=3000 K.
Two shells at 1000K and 400K are confirmed in the IR SED.\\

\centerline{\bf RR Tel}

Also the IR SED of RR Tel is well reproduced by emission from dust shells
at 1000 and 400 K. The temperature of the giant is of 2450 K.
The nebulae were modelled by Contini \& Formiggini (1999, Table 1)
 with models rev1, exp, and rev$_{XX}$, as reported also in Table \ref{tab:suma}.

The emission from the WD is emerging in the UV frequency range and is reproduced by a bb
corresponding to T= 1.4 10$^5$ K, in agreement with the temperature adopted to
explain the line spectra.

RR Tel modelling reveals also that the shock front facing the
red giant, which is  generally negligible in other SSs, is
contributing to the spectra, while the contribution of the expanding shock 
in the extended external region  is in this case not relevant.
The two component (reverse and expanding) which contribute
to the lines are hardly recognizable in the SED  because the shock velocities
are very similar.
Model rev$_{XX}$ which explains the X-ray is very similar to those found in Paper I
 by fitting the IR line ratios.\\

\centerline{\bf V627Cas}

There is only one datum in the radio, an upper limit which  constrains 
the models. The radii of the nebulae thus are also  upper limits. We used the reverse shock model from BI Cru.
The temperature of the Mira is approximated by T=2450 K, while two main shells of 
dust at T=1000 K and 400 K are recognizable in the IR.\\

\centerline{\bf R Aqr}

This SS was modelled in details by Contini \& Formiggini (2003),
therefore we take their models  to explain the data.
The Mira temperature is approximated by T=2500 K.
Interestingly, this is the only case where just one dust shell (at T=1000K) is evident from the IR data. 
The  data are not enough  to  permit to understand  whether a cooler shell is actually present, 
or if R Aqr is a peculiar system under this aspect.\\

\section{Discussion}
A consistent understanding of the  symbiotic phenomenon has been possible only after their binary 
nature was acknowledged. 
Moreover,  the study of dust grains and, in general, of dust shell properties in SSs is interesting 
not only in itself, but also in the light of SSs as colliding-wind systems, namely systems where shocks are at work. 
We thus have the unique opportunity of studying in a relatively small region both dust condensation and 
destruction processes, as well as the mutual interaction of grains with a very dynamic ambient gas.  
The results of dust features in SSs are given in a previous paper (Paper I).

After having briefly described the single objects one by one, we would like to investigate some common properties 
that appear evident in our analysis.  
 This can be done because  we can apply to the continuum SED from the nebulae, the solutions 
of the   calculations used for  modelling  the line spectra. They are unique because the
line ratios  are strongly constraining. The  models are further cross-checked
 comparing the calculated fluxes with the observed absolute fluxes.
This leads, for instance, to the calculation of the radius of the nebulae which must 
be compatible with
the geometrical dimensions of the system (e.g. the binary separation for the reverse shock, etc).

There  is strong evidence  that  virtually all SSs of the sample have at least two dust 
shells, at about  "standard" temperatures.  
It can be asked whether  are there  two dust shells or dust with a range of temperatures.
Considering single AGB stars, it seems that there is a monotonic continuity  of grain temperature profiles, starting
from the highest value defined by the dust temperature at the condensation radius. We  could refer more properly
to dust within a temperature range.
In SSs, on the other side, the colliding-wind binary nature of the system causes a real discontinuity in the dust
spatial distribution (e.g. more than a single dust shell) since the WD radiation field and shocks modify
the local physical conditions, in some cases even preventing (Paper I) or altering (Angeloni et al. 2007c)
the dust condensation sequence.

Two peaks are generally directly recognizable from the SED. Other peaks
between them,  corresponding to intermediate temperatures, would be less
intense in order to
be unseen. Lower fluxes would  correspond to  radii  smaller than that of
the  hot dust shell. This is  unsuitable to a smooth distribution of dust temperatures with radii.

Actually the theoretical models predict the  shell
temperatures and radii by the direct comparison with the SED
observational data.
It can be noticed from the diagrams in Fig. 4,  that the dust thermal emission  emerges from
the SED profile  as an infrared excess easily  recognizable as a  black body like bump 
corresponding to T$\sim$1000 K.
Moreover,  a cooler dust shell  (T$\sim$ 400 K)  often  appears, while dust reprocessed radiation
from the shocked nebulae seldom emerges over the bremsstrahlung.  

The shell radii  are obtained by  adjusting the results of consistent modelling (Table \ref{tab:res}) 
to the continuum {\it absolute} fluxes observed at Earth.
Comparing  dust shell sizes with 
the most likely values reported in the literature for symbiotic binary separation, 
it seems that the hottest  (T$\sim$ 1000K) dust shells are the "classic" circumstellar 
shells known to surround Mira 
and other AGB stars, while the coolest (T$\leq$ 400K) ones  are likely to be circumbinary structures, 
namely they surround the whole system. This seems to be a peculiar feature of D-type systems.

The 1000 K shell would always be distinguishable, because it is actually an expanding dusty
envelope somehow "detached" from the star itself: in fact, the suitable physical and chemical
conditions for  grain formation are usually reached  beyond a few stellar radii (see e.g. Danchi et al. 1995).
Moreover, we have tried to demonstrate  that the coolest dust shell is mainly circumbinary,
and clearly recognizable on the SED only if the dust grains survive  sputtering and evaporation.

In the following, we  discuss some further correlations  which highlight specific 
features of dust shells in SSs, e.g.  the luminosity of symbiotic dust shells, the links between 
temperature of dust and radius of the corresponding shell, also in comparison with a sample of other dusty stars. 
Eventually, we have looked for any relations between dust shell radii and orbital periods.

\begin{table}
\centering \caption{The radius of nebulae and shells \label{tab:res}}
\begin{tabular}{lllllllll}\\ \hline  \hline\\
\ object & T$_{Mira}$& r$_{rev}$  &  r$_{exp}$  & r$_{1000}$ & r$_{400}$\\
\hline
\ BI Cru & 3000 & - &7 & 1.5e14$^1$   &1.6e15$^2$ \\ 
\ SS73 38$^3$  & 3000 & 6.3e13  &3.5e15 &1.5e14  &2.2e14  \\
\ V835 Cen & 3000 & 3.4e14    &8.9e15  &7.9e13& 2.5e14  \\
\ H1-36  & 2500 &1.e14  &  -     & 4.4e14$^1$   & 4.7e15$^2$\\
\ HM Sge &3000&7.7e14    & 6.9e15 & 2.45e14& 1.1e15   \\
\ V1016 Cyg  & 3000 & 6.22e14     &3.9e16 &2.5e14  & 8.8e14  \\
\ RR Tel & 2450 & 1.36e15    &1.1e15 & 3.4e14 & 8.6e14\\ 
\ V627 Cas & 2450 & 1.9e14 &1.2e16 & 1.5e14 & 5.37e14 \\             
\ R Aqr  & 2500 & 6.e14      &6.e15   &1.9e14  & -  \\
\hline
\end{tabular}
\flushleft
$^1$ calculated by a bb of 800 K;\\
$^2$ calculated by a bb of 250 K;\\
$^3$ calculated with distance from Earth of  4.8 kpc
\end{table}

\subsection{Dust shell luminosity}

\begin{figure}
\begin{center}
\includegraphics[width=0.45\textwidth]{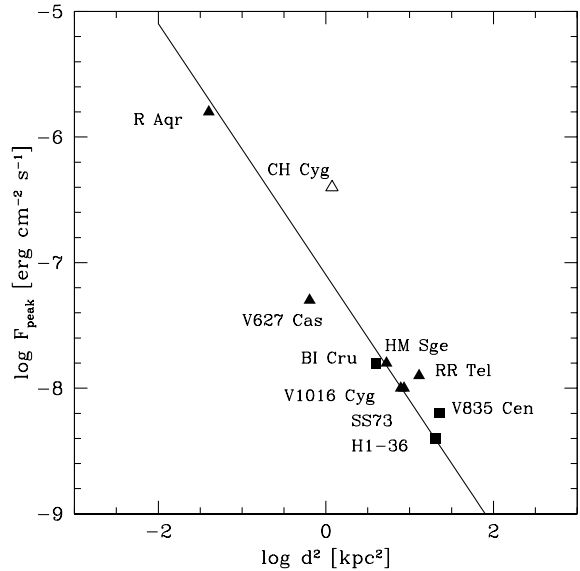}   
\end{center}
\caption{The flux at the peak of the dust shell closest to center  as a function
of the distance from Earth of SSs. Black triangles: for a dust temperature of 1000 K.
Black squares: for a dust temperature of 800 K. Open triangles: for a dust
temperature of 1000 K in the CH Cyg system during the quiescence phase 1996-97.
\label{fig:lum}}
\end{figure}

We consider  the flux at the peak (F$_{peak}$) of the bb emission,
corresponding to the dust shell closest to the Mira, at the highest temperature.
In Fig. \ref{fig:lum}  F$_{peak}$  vs.  the square distance to Earth
 is shown for all the objects of the sample.
Most of the sample objects have a shell corresponding to 1000 K.
We have included also the objects with a shell at 800 K (BI Cru) and CH Cyg,
which is not definitively classified as a D-type (see however Contini et al. 2009c),
for comparison. The data are  taken from the diagrams in Fig. \ref{fig:seds} and Table \ref{tab:res}.
Recall that $F\propto$ $L$/$d^2$, where $F$ is the flux, $L$ the luminosity,
and $d$ the distance from Earth. 
The objects, except  V835 Cen,  follow as expected a straight line because the luminosity
mainly depends on the temperature of the emitting shell.
Moreover, the dusty material
had not  yet the time to merge with the interstellar matter.
The line in the diagram is defined by HM Sge, V1016 Cyg, SS73 and R Aqr.

We employ this diagram  to guess out a likely distance from Earth of SS73 38.
In fact,  from the diagram in Fig.  \ref{fig:seds} we measure a flux of 10$^{-8.2}$ \erg
at the peak of the 1000 K shell. The point at this flux on the line
in Fig. \ref{fig:lum} should correspond to a distance of $\sim$ 3.6 kpc  slightly less than the 4.8 kpc given by 
Gromadzki et al. (2009), and derived exploiting the period-colour relation for Miras as in Whitelock et al. (2006).
Moreover, for V835Cen, Whitelock (1987) gives a distance of 2.8 kpc 
in agreement with Fig. 4 diagram, while the 9.4 kpc distance by Feast et al. (1983) is clearly over predicted.

\subsection{Dust shell radius}
In Fig. \ref{fig:danchi} we have plotted the radius of the shells
versus the dust temperature. Comparing with the data of Danchi et al. (1994) for isolated stars,
the  temperatures  of symbiotic dust shells reach similar values.
However, the  dust shell temperatures
regarding the sample
are  generally  concentrated at $\sim$ 1000 K and $\sim$ 400 K, while in the case of
late giants, they spread
more uniformly throughout the same range. The origin of the gap can be
ascribed to the WD radiation which can lead to evaporation of the grains in its
proximity.

\subsection{Orbital periods}

Unfortunately, reliable values for symbiotic orbital parameters are known only for S-types. Some further speculations 
have been proposed also for the quite rare D'-type objects (Zamanov et al. 2006), but no definite results exist for D-types.
For these dusty objects, there are just rough constraints about binary separation based on dynamical considerations 
(i.e. hosting a Mira giant forces the system to have an orbital period of at least some decades of years) and, 
in any case, the uncertainties in several basic parameters are too large.

Nonetheless, we would like to  investigate how the dust shells periodically ejected by the 
Mira feel the influence of the compact, much hotter object. 
Presumably, the very circumstellar region around the WD is dust-free, because 
grains can not survive such high temperatures  (Munari et al. 1992). 
In contrast, the region of the Mira facing the center of the system might not be so 
disturbed by the diluted WD radiation field as to  prevent the condensation of 
dust grains, because of the large binary separation. 

Another aspect that  must be taken into account is the turbulent character of the 
interbinary colliding-wind regions, which leads to  sputtering 
rather  than evaporation, as the most important grain destruction mechanism 
(Paper I). 

\begin{figure}
\begin{center}
\includegraphics[width=0.45\textwidth]{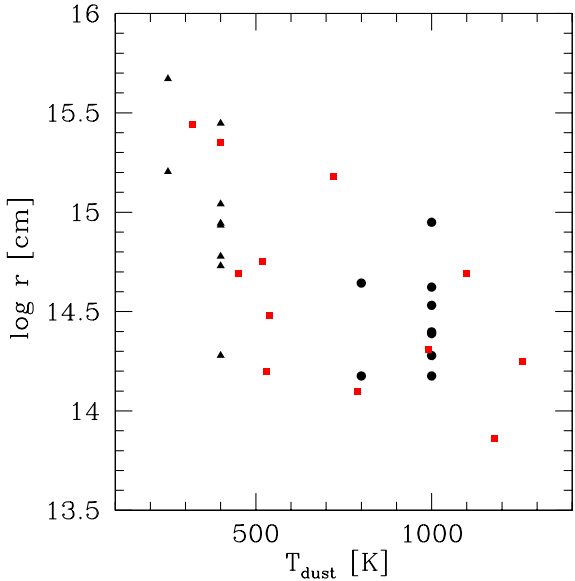}       
\end{center}
\caption{
Radius vs. temperature for the dust shells. Squares: data from Danchi et al. 1994; black circles: 
symbiotic dust shells at the highest temperatures ($\sim$ 1000 K); black triangles: symbiotic dust shells at 
the lowest temperatures ($\leq$ 400 K).\label{fig:danchi}}

\end{figure}

\begin{figure}
\begin{center}
\includegraphics[width=0.45\textwidth]{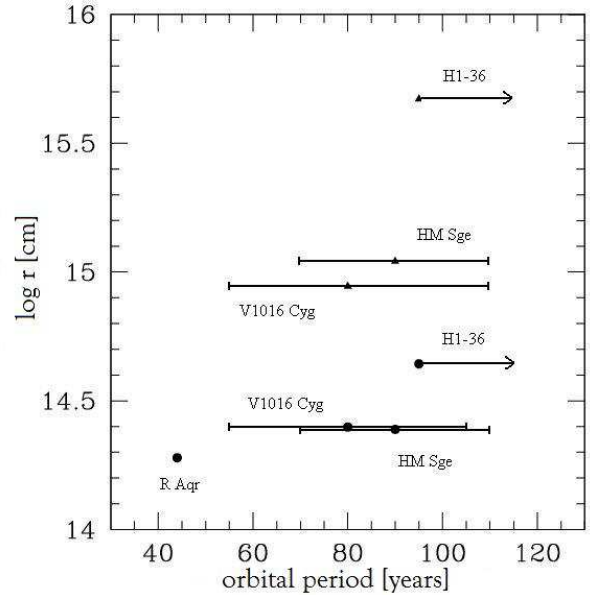}    
\end{center}
\caption{Dust shell radii vs orbital periods. Circles: shells at 1000K; triangles: shells at 400K.\label{fig:orb}}
\end{figure}

In order to find out if the symbiotic character influences also the dust shell 
parameters, we  investigate the correlation between 
the dust shell radii, as calculated by the SED modelling, and the orbital periods as 
reported in literature for some objects of our sample. Fig. \ref{fig:orb} shows 
that, within the error bars, a significant trend is recognizable: \textit{the 
longer the orbital period, the larger the shell radii. }
Even though these results cannot be quantifi�ed yet (our sample is not statistically 
significant and uncertainties in several basic parameters are too large), 
this correlation looks meaningful enough to encourage further investigations.
Eventually, a last consideration is common to Figs. \ref{fig:lum}, \ref{fig:danchi} 
and \ref{fig:orb}. The
representing points of the dust shell radius lie in two distinct regions of the 
diagrams: the
first one, at log r $\leq$ 14.7, populated by the hottest shells, and the other one at log r $>$ 15, by
the coolest  ones. Interestingly, the gap in the shell radius distribution overlaps the median
strip corresponding to the typical binary separation for D-type SS. This indicates that  dust grains
cannot condensate, or easily evaporates, along the WD orbit where the radiation field of
the hot component is too harsh for grain survival.

The presence of a companion star splits the system into two regions, an interbinary
one and a circumbinary. The interbinary region is filled by the shell surrounding the Mira
and expands with a velocity of about 30 \kms , hence reaching a distance of about 10$^{14}$ cm
in a very short time ($\sim$ 1 year) with respect to the orbital period, and then  expands outwards. Close
to the WD the dust grains are easily evaporated, hence causing the gap in the distribution
of dust throughout the system. This is reflected by the distribution of the shell radii in
Fig. \ref{fig:danchi}. When the shell reaches a distance of 10$^{15}$ cm, its mass, calculated by the mass loss rate (m$_{shell}$ $\sim$  10$^{-6}$ \msol 
assuming \.{m}=10$^{-7}$ \msol/y), is of the same order of the mass swept 
up from the circumstellar
region  (m$_{swept}$ $\sim$ 1.3 10$^{-6}$ \msol, adopting a shell geometrical thickness
of 10$^{13}$ cm and a gas density of 10$^7$ \cm3), therefore, the shell  will lose any 
identity by merging with
the interstellar medium. This could explain why  just one external shell is recognizable. 
New spectra in the extreme far-IR,  allowing to detect cooler and cooler
dust,  will reveal  other outer structures before  merging  with the ISM.

\section{Concluding remarks}

We have collected continuum data in the IR 
for a sample of D-type SSs.
For most of the objects the detailed modelling
of the line spectra and the continuum SED in previous specific
works leads to an accurate description of the physical characteristics
and of the morphological structure throughout each system.
For the other objects we have guessed the constraining parameters
from a   first analysis of the line ratios and/or by similarity
with other well-known objects.

The study of dust grains  and, in general, of dust shell properties in SSs is interesting 
not only in itself, but also in the light of SSs as colliding-wind systems, namely systems where shocks are at work. 
We thus have the unique opportunity of studying in a relatively small regions both dust condensation and 
destruction processes, as well as the mutual interaction of grains with a very dynamic ambient gas. 

By modelling the SED continuum for each object,
we have found  the  common characteristics to all the systems:
1) at least two dust shells are clearly present at about 1000 K and 400 K,
2) the radio data are explained by  thermal self-absorbed emission from the reverse
shock between the stars, while 3)  the data in  the  long wavelength tail
come from the expanding shock outwards the system.
4) In some system the contribution from the WD in the UV can be directly seen. Finally, 5) soft X-ray produced by bremsstrahlung from the reverse-shock nebula
between the stars, are predicted for some objects.

The results thus confirm the validity of the colliding wind model
and the important role of the shocks.
The dust shells which were so important in explaining some features
in the light curve at different phases of the  CH Cyg	
 (Contini et al. 2009c)
are well recognizable in all the D-type objects.
We wonder whether their presence will have an impact
on the IR light curve at certain phases.

The flux peak of the 1000 K dust shells are inversely correlated
with the square distance to Earth. This permitted  to
guess out the distance of SS73 38 and to select the appropriate distance of V835 Cen.

By consistent modelling of the dust shell at 1000 K in R Aqr, we have found 
that the dust-to-gas ratio by mass 
 is $\geq$ 0.1, even higher than that evaluated for 
 supernovae.

The comparison of the flux calculated at the nebula with those observed
at Earth   reveals the  distribution throughout the system
of the
different components, both nebulae and dust shells.
The correlation of  shell radii with the orbital period show that  larger radii
are found at larger periods. 
Moreover, the temperatures of the dust shells 
regarding the sample,
are  generally  concentrated at $\sim$ 1000 K and $\sim$ 400 K, while in the case of
late giants, they spread
more uniformly throughout the same range
because the WD radiation most probably  leads to evaporation of the grains in its
proximity.

\section*{Acknowledgments}
We are very grateful to the referee for constructive criticism and helpful comments.
We thank E. Leibowitz for interesting conversations.
RA remembers M. Mariani for his constant encouragement during the very preliminary version of this work.

\end{document}